\documentclass[11pt]{article}
\usepackage{epsf,fancyheadings,amsfonts,amssymb}
\usepackage[centertags]{amsmath}

\newcommand{\beqn}{\begin{eqnarray}}
\newcommand{\eeqn}{\end{eqnarray}}
\newcommand{\beqnst}{\begin{eqnarray*}}
\newcommand{\eeqnst}{\end{eqnarray*}}
\newcommand{\bs}{\boldsymbol}

\begin{document}

\begin{titlepage}
\begin{flushright}
IMSc/2003/01/01 \\
\end{flushright}
\begin{center}
{\large {\bf {Classical and Quantum Mechanics of Anyons}}}
\end{center}
\vspace{1cm}
\begin{center}
G. Date*,  M.V.N. Murthy* and Radhika Vathsan$^\dagger$
\end{center}
\begin{center}
*The Institute of Mathematical Sciences, Madras 600 113, India\\
$^\dagger$ Birla Institute of Technology and Science, Pilani (Rajasthan) 333 031
\end{center}
\vspace{2cm}
\begin{abstract}

We review aspects of classical and quantum mechanics of many anyons
confined in an oscillator potential.  The quantum mechanics of many anyons
is complicated due to the occurrence of multivalued wavefunctions.
Nevertheless there exists, for arbitrary number of anyons, a subset of
exact solutions which may be interpreted as the breathing modes or
equivalently collective modes of the full system.  Choosing the
three-anyon system as an example, we also discuss the anatomy of the so
called ``missing'' states which are in fact known numerically and are set
apart from the known exact states by their nonlinear dependence on the
statistical parameter in the spectrum. 

Though classically the equations of motion remains unchanged in the
presence of the statistical interaction, the system is non-integrable
because the configuration space is now multiply connected.  In fact we show that
even though the number of constants of motion is the same as the number of
degrees of freedom the system is in general not integrable via
action-angle variables. This is probably the first known example of a many body
pseudo-integrable system. We discuss the classification of the orbits and
the symmetry reduction due to the interaction. We also sketch the
application of periodic orbit theory (POT) to many anyon systems and show
the presence of eigenvalues that are potentially non-linear as a function
of the statistical parameter. Finally we perform the semiclassical
analysis of the ground state by minimizing the Hamiltonian with fixed
angular momentum and further minimization over the quantized values of the
angular momentum.

\end{abstract}
\end{titlepage}

\tableofcontents
\newpage
\section{Introduction}

In 1977, Leinaas and Myrheim\cite{lm} showed that in two space dimensions
it is possible to have particles obeying intermediate statistics different
from the well known Bose and Fermi statistics.  Wilczek\cite{wilczek}
later coined the name {\it anyons} for particles obeying these peculiar
statistics. In the last two decades, a lot of work has been done in this
field. Many of these developments have been brought up to date and nicely
summarized in the books by Lerda\cite{lerda} and Khare\cite{khare} as also
in the collection of articles on anyon superconductivity\cite{supercond}. 
One aspect that has not been covered in detail in these books and other
reviews\cite{reviews} concerns the question of integrability properties of
a system of many anyons and its classification. This review is intended to
focus on these aspects which have not been discussed in detail earlier or
have been largely left out of earlier attempts. 

Here we focus on the non-relativistic classical and quantum mechanics of
particles in two spatial dimensions which obey fractional statistics
\cite{lm,wilczek}. Generically such particles are referred to as anyons. 
Anyons as a research area is now over twenty years old. These systems have
emerged as being interesting in their own right from the point of view of
mathematical physics at both the classical and the quantum level.  These
systems also constitute an example of inequivalent quantizations due to
non trivial fundamental group of the configuration space \cite{cmp}.  It
is well known that the existence of fractional statistics is intimately
connected to having multivalued wavefunctions which naturally occur in
quantum mechanics on multiply connected spaces \cite{braid}.  The
classification of multivalued wave functions is provided by one
dimensional representations of the fundamental group of the multiply
connected configuration space.  The configuration space $Q^N$ is the
$d$-dimensional Euclidean space given by $R_2^N$ with all the diagonal
points, $\Delta$, removed.  The fundamental group of $Q^N$ turns out to be
$$\pi_1(Q^N) = S_N ~~if ~~d \ge 3$$ and $$\pi_1(Q^N) = B_N ~~if ~~d =2 $$
where $B_N$ is the braid group of N objects which contains the permutation
group $S_N$ as a subgroup. This immediately brings out the difference
between two and three dimensions.  Although there are several known
examples of a kinematic classification \cite{cmp} (in terms of the
representations of the fundamental group), very little is known about the
dynamics on such spaces.  The motivation for studying such systems is 
two fold: they are a good and physically relevant example of quantum
mechanics on multiply connected spaces.  It is physically relevant because
anyons were proposed as candidates for explaining fractional quantum Hall
effect \cite{fqh} and their possible connection to high $T_c$
superconductivity \cite{htc}.  It is a good example because at least some
exact solutions to the energy eigenvalue problem for N anyons in some
external confining potential are known---an exception in many-body quantum
mechanics with non-separable Hamiltonians.

The quantum spectrum of anyons has been analyzed by many authors.  Broadly
speaking, several authors have pointed out the existence of a subset of
exact solutions\cite{exact} whereas in the numerical analysis of the low
lying spectrum of three and four anyons, many non-trivial features of the
spectrum have been shown to exist\cite{numeral}. The quantum mechanical 
spectrum (with
harmonic potential added) shows two distinct qualitative features: a)
eigenvalues  which depend {\em{linearly}} on the statistical
parameter, $\alpha$, {\it all} of which are exactly known and b) eigenvalues
which depend {\em{non-linearly}} on $\alpha$ and {\it none} of which is exactly
known\cite{numeral}. This is also borne out by a perturbative analysis of
the three-anyon ground state\cite{mccabe,perturb}. Another interesting 
feature of
the spectrum is the additional degeneracy around the semion point which is
related to the reflection symmetry\cite{sen}.  These features may be
traced to the properties of partial separability and pseudo-integrability
manifested at the classical level \cite{pseudo}. 

Partial separability of the Hamiltonian may be exhibited explicitly in
terms of two collective degrees of freedom and the remaining ``relative"
degrees of freedom. The total Hamiltonian, after removing the center of
mass, may be written as a sum of $H_1(collective) + H_2(collective,
relative)$. The commutator of $H_1$ and $H_2$ turns out to be proportional
to $H_2$ \cite{pseudo}. This implies that the subspace of eigenstates of 
the full $H$ on which $H_2$ vanishes are exact eigenstates of $H_1$ and 
these give {\em{all}} the exactly known eigenvalues and eigenstates. 

Pseudo-integrability, a concept first introduced by Richens and Berry
\cite{berry}, may be described by exhibiting 2N constants of motion in
involution, which fail to lead to integrability via usual action-angle
variables. This was conjectured to be the reason for the level repulsion
seen numerically in the non-linearly interpolating spectrum \cite{pseudo}.
The precise meaning of pseudo-integrability in this context, however, was
not elaborated earlier. In particular, how this classical feature
translates into there being only two good quantum numbers, was not
analyzed. A further fallout of the pseudo-integrability property of the
anyon system is its effect on the so called nearest neighbour spacing
distribution in the energy spectrum. It is now known that the spectrum
exhibits features of both regular and irregular spectra as a function of
the statistical parameter\cite{djm}. 

Yet another method of approaching the quantum spectrum of many anyons is
through the stationary phase approximation (SPA). In this, an approximation
to the propagator $G(E + i \epsilon)$ is developed using a suitable path
integral representation \cite{semiclassical}. In this approach the
propagator is typically obtained as a sum over (families of) periodic
trajectories in the classical phase space.  This is entirely given in
terms of classical quantities. One therefore expects to see directly the
effects of non-trivial fundamental group of the phase space,
pseudo-integrability and partial separability in a semiclassical
framework.  Since at the classical level, anyons with oscillator
confinement, are {\it locally} identical to the oscillator system (but
differ in the global topology of the phase space), the periodic orbits are
known. Thus application of the periodic orbit theory (POT) to anyons, may
be expected to be tractable. 

There is one clarification to be noted. At the quantum
level, the statistical parameter, $\alpha_q$, enters via the stipulation
of multivaluedness of the wave functions and is {\it dimensionless}. At
the classical level its counterpart, $\alpha_c$, enters as the coefficient
of a total derivative term in the Lagrangian and has the dimensions of
$\hbar$. These two must be related as $\alpha_c = \hbar \alpha_q$. In the
SPA computation, $\alpha_c$ is held fixed with $\hbar$ going to zero. In
the leading approximation, the spectrum will depend {\it linearly} on
$\alpha_c$ and thus also on $\alpha_q$.  Alternatively, for comparison
with the quantum spectrum to leading order, one will want to keep
$\alpha_q$ fixed implying that $\alpha_c$ to go to zero with $\hbar$.
Viewed either way, one may {\it not} expect to see {\it non-linear}
dependence on the statistical parameter at the level of the leading
approximation. Nonetheless, if the non-linearly interpolating eigenvalues
also have a linear piece, one should be able to see it in the leading SPA. 

This article is concerned with the discussion of the many points raised
above. While many of these points have emerged from previously published
papers, we attempt a coherent presentation of all these aspects with
appropriate connections noted along with new material. We 
begin in section 2 with the basic
formalism where we build the Hamiltonian of the anyons in two dimensions
by requiring that their exchange statistics depend on a continuous
parameter. In section 3, we review the quantum spectrum of many anyons. We
first prove the existence of a subset of exact solutions. We then discuss,
albeit in the case of three anyons, the so called missing states.  Noting
that eigenvalues themselves are independent of where the differential
equation is solved, we expect the asymptotic form of the differential
equation to determine the spectrum.  The three anyon problem in the
asymptotic regions reduces to solving a Confluent Hypergeometric Equation
(CHE) in the region $R \le x < \infty $.  This equation has a solution
regular at $x=0$ and a solution which is irregular at $x=0$.  The regular
solution leads to linearly interpolating energies while the irregular one
may (R is strictly greater than zero) lead to nonlinearly interpolating
solutions.  Though we are unable to make definite predictions about the
spectrum, we point out the inherent difficulty which involves the question
``what is the quantization condition''. 

In section 4, we analyze the many anyon system classically and show that 
the subset of exactly solvable solutions arise
from the quantization of the collective coordinates (after the trivial
centre of mass degrees are separated).  This is made possible by the
property of partial separability of the system for certain initial
conditions and consistent with the equations of motion. In fact these
solutions do not carry any information on the internal dynamics of anyons
which may be frozen as far these solutions are concerned.  These
eigenvalues also goes as $N^2$ for N anyons.  Thus the solutions are trivial
dynamically. The real anyon dynamics therefore resides in the solutions
which depend non-linearly on the statistical parameter.  In the analysis
of the quantum virial expansion for the equation of state of the anyon
gas, it was shown that these trivial states conspire to cancel the
divergent parts while the nonlinear solutions give the finite part which
then defines the equation of state\cite{eqst}.  Thus it is of utmost
importance to obtain these non-linear states to study the thermodynamic
properties of the system. The low lying states have been obtained
numerically\cite{numeral} and perturbatively \cite{mccabe,perturb} in the 
case of
three and four anyons. In the case of large N anyonic systems,
Thomas-Fermi approach has been used to obtain the properties of the ground
and low lying excited states\cite{tf}. However to establish a virial
expansion for the equation of state, by going beyond the third
and the fourth virial coefficients, much more work on the {\it 
nonlinear} states is needed. 

In section 5, we discuss semiclassical approach via the periodic orbit
theory.  we consider the question of classical integrability- the
equations of motion of course do not depend on the statistical parameter,
however the configuration space of N- particles is $R_2^{N} - \Delta$,
where $\Delta$ denotes the set of all diagonal points which are removed. 
All the orbits which pass through these diagonal points are now modified.
This leads us to conjecture that the system may be non-integrable a-la
Richens and Berry \cite{berry}. We demonstrate that the classical analysis
is adequate for exhibiting quantum symmetries. In this process we sharpen
and clarify the meaning of pseudo-integrability and role of the
fundamental group of the phase space.  The system is shown to be identical
to the isotropic oscillator {\em{locally}} but not at the global
topological level. We then classify the classical trajectories using
symmetry transformations and explain how the non-trivial global topology
reduces the symmetry group.  Furthermore, we sketch the application of the
periodic orbit theory to many anyons. In the process we show the presence
of eigenvalues which are potentially non-linearly interpolating. As
mentioned earlier such eigenvalues have been seen in the numerical
spectrum for three and four anyons \cite{numeral} and have been
conjectured to be present for general $N$ \cite{numerical}. We obtain
these systematically from POT.  Further, both the classical modeling and
the exact propagator for two anyons, indicate an ambiguity regarding
possibility of closed orbits with half the basic period and its inclusion
in the trace formula. We discuss this ambiguity in some detail. We show
that it is possible to regularise the point-like statistical flux and
deduce the existence of half period trajectories. We look for their
evidence in the exact propagator for two anyons and point out their
ambiguous status.  In Section 6, we present a detailed semiclassical
analysis of the ground state of many anyons using a two step minimization.
While this may seem unusual procedure to follow, we show that the method
(even at the classical level) retains some of the features of the quantum
spectrum. 

Finally the results are summarized in Section 7. The appendices contain
many results to keep this review self contained.  Appendix A contains a 
discussion of regularisation method for the so-called reflecting orbits 
which arise in the zero angular momentum sector.  Appendix B contains
details of $OSp(4N, R)$ symmetry.  Appendix C  contains details of the 
Symplectic diagonalisation procedure relevant for Section 6.

\section{Basic Formalism}

In this section, we first develop the concept of anyons in two dimensions
through multivalued wavefunctions.  The exchange statistics of anyons then
involves a continuous parameter, $\alpha$---the so called statistical
parameter\footnote{ Later we replace $\alpha$ with $\alpha_q$ to denote
the fact that this is the statistical parameter appearing in the quantum
mechanics of anyons. While this is a dimensionless parameter here, it is
not so in the classical case. This distinction is important in the
discussion of semiclassics.}.  It is the peculiarity of two dimensions that
the quantization does not depend of the particular value of $\alpha$,
unlike in higher dimensions where we require $\alpha$ to be an integer. 

In the following, by ``anyons'' we mean a quantum mechanical system of N
particles in two dimensions with wave functions which have a
stipulated multi-valuedness to be specified below.  To make this explicit
let us denote a generic wave function as $\psi (\vec{r}_1,...,\vec{r}_N)$,
where $\vec{r}_i$ denotes the position vector of a particle.  Let
$[P_{ij}]_{\gamma}$ denote the operation of taking the $i$th particle
coordinate around the $j$th coordinate along a closed path $\gamma$.  The
path $\gamma$ does not enclose any other particle coordinate and is taken
in an anti-clockwise sense, say.  Then let us stipulate that under such an 
operation $\psi$ acquires a phase namely,
\begin{equation}
[P_{ij}]_{\gamma}\psi (\vec{r}_1,...\vec{r}_N) 
= exp(i2\pi\alpha)\psi (\vec{r}_1,...\vec{r}_N) .\label{eq2.1}
\end{equation}
If a path $\gamma$ encloses other particle coordinates as well then
such a path can be broken into a set of closed paths each of which
encloses exactly one particle.  Applying the stipulation
above, one can compute the total phase, for such a path.  If the sense
of the path is reversed then $\alpha \rightarrow -\alpha$.  Clearly
the phase acquired depends only on the homotopy class of the 
path (i.e., it is the same for two paths $\gamma$ and $\gamma '$ if 
$\gamma$ and $\gamma '$ can be continuously deformed into each other).

Let us introduce the complex notation for particle coordinates:
$z_j = x_j +i y_j, ~ \bar z_j = x_j -i y_j$.  
Clearly $z_{ij}$, where $z_{ij}=z_i - z_j$, has the
property that if $z_j$ is taken around $z_i$, $~z_{ij}^\alpha$ changes
by  $\exp(i2\pi\alpha)$.  This allows us to write any generic wave
function satisfying eq.(\ref{eq2.1}) as,
\begin{equation}
\psi (z_i,\bar z_i) = \left [ \prod_{i<j}\left(\frac {z_{ij}}{\bar
z_{ij}}\right)^
{\alpha/2} \right ]\tilde\psi (z_i,\bar z_i),
\end{equation}
with the bracketed expression being a phase
and now  $\tilde \psi (z_i,\bar z_i)$ is a single valued function. 

Clearly, 
\begin{equation}
\nabla_k \psi (z_i,\bar z_i) = \prod_{i<j}\left(\frac {z_{ij}}{\bar
z_{ij}}\right)^
{\alpha/2} \left [ \nabla_k \tilde\psi (z_i,\bar z_i) 
+ \nabla_k ln\left(\prod_{i<j}\left(\frac {z_{ij}}{\bar z_{ij}}\right)^
{\alpha/2}\right) \tilde\psi (z_i,\bar z_i) \right ].
\end{equation}
which can be rewritten as,
\begin{equation}
\nabla_k \psi (z_i,\bar z_i) = \prod_{i<j}\left(\frac {z_{ij}}{\bar
z_{ij}}\right)^
{\alpha/2} \left [ \nabla_k \tilde\psi (z_i,\bar z_i) 
+ i\alpha \sum_{j\ne k}\frac{\hat z \times \vec {r}_{kj}}
{{\mid \vec {r}_{kj} \mid}^2} \tilde\psi (z_i,\bar z_i) \right ].
\end{equation}
Since $\tilde \psi$ is single valued, the right hand side  of the above 
equation has 
exactly the same multivaluedness as the left hand side.  In other words we 
have, 
\begin{equation}
\nabla_k \left[\prod_{i<j}\left(\frac {z_{ij}}{\bar z_{ij}}\right)^
{\alpha/2} \tilde\psi (z_i,\bar z_i)\right] =
\left[\prod_{i<j}\left(\frac {z_{ij}}{\bar
z_{ij}}\right)^\frac{\alpha}{2}\right] D_k \tilde \psi,
\end{equation}
where
$$ D_k \tilde \psi = [\nabla_k + \vec A_k ]\tilde \psi $$
and
$$ A_k(\vec r_k) = i\alpha \sum_{j\ne k}
\frac{\hat z \times \vec {r}_{kj}}
{{\mid \vec {r}_{kj} \mid}^2}. $$

This allows us to write any higher order differential operators on
$\psi$ in terms of corresponding covariant differential operators on
the single-valued wave function $\tilde \psi$.  In particular a
Hamiltonian operator, typically $ - \sum_i \nabla^2_i +V$ can be 
written similarly.  An eigenvalue equation written in terms of 
$\psi$ can then be recast as a corresponding equation in terms of
$\tilde \psi$ involving the covariant derivative.  

Although both formulations are equivalent, dealing with operators
on multivalued wave functions is much less transparent than 
dealing with operators on single-valued wave functions.  Naive 
commutation rules, symmetries that one would expect by looking 
at an operator on single valued functions are not at all true 
in general for the ``same" differential operators acting on multivalued
wave function.

Considering eigenvalue problem in terms of $\tilde \psi$ has other
advantages too.  Since all the subtleties of multivaluedness are
equivalently transcribed in terms of additional ``interaction''
terms (the so called statistical interactions), the eigenvalue
problem is amenable to approximations.  One is also on firmer ground 
in doing usual algebraic manipulations with operators.   With these in 
mind we will work with single-valued wave functions with ``statistical
interactions''.  

As a first step one would like to understand the system of ``free 
anyons''.  However, the statistical interaction falls off as
$\mid r_{ij}\mid ^{-2}$ as $\mid r_{ij} \mid \rightarrow \infty$.
So one is not sure whether the Hamiltonian with only statistical
interactions has only discrete eigenvalues.  One can put the 
system in a box to ensure discrete eigenvalues but then one needs
suitable boundary condition. An oscillator potential ensures 
discrete spectrum without introducing a finite size.  One could
take some other confining potential but in the limit $\alpha \rightarrow
0$ one should know the spectrum. One then has hope of doing at least
the perturbative analysis \cite{ouvry}.  Since the statistical interaction 
depends only on relative separations, the Centre of Mass (CM) dynamics 
should play a trivial role and oscillator potential also allows
a separation of CM and relative coordinate dynamics.  The oscillator 
potential problem can also be mapped on to a problem of anyons in
a real, constant external magnetic field along the z-axis \cite{hotta}.  
Bearing these facts in mind, we choose the oscillator potential
without further justification.  In order to derive the thermodynamic
properties of a system of anyons there exist well defined methods
of eliminating the dependence on the oscillator frequency
\cite{mccabe,eqst}.

The Hamiltonian we consider is -- after carrying out the usual scaling
of variables 
-- in terms of dimensionless quantities
  
\begin{equation}
H =\hbar \omega[\frac {1}{2}\sum_{i=1}^{N}{p}_{i}^{2} +
\frac {1}{2}\sum_{i=1}^{N} r_{i}^{2}
-\alpha\sum_{j>i=1}^{N}\frac{\ell_{ij}}{r_{ij}^2}
+\frac{\alpha^{2}}{2}\sum_{i\neq j,k}^N\frac{\vec{r}_{ij}.\vec{r}_{ik}}
{r_{ij}^{2}  r_{ik}^{2}}], \label{anyham}
\end{equation}
where
\[\ell_{ij}= (\vec{r}_{i} - \vec{r}_{j})\times(\vec{p}_{i}-\vec{p}_{j}).\]
and all distances have been expressed in units of
$\sqrt{\frac{\hbar}{m \omega}}$.
Notice that the statistical interaction is independent of the centre of
mass coordinates.
This is the Hamiltonian we analyze in the following sections both quantum 
mechanically and classically. In the quantum analysis, the Hamiltonian
is considered to act on wave functions which vanish suitably at the
coincident points.


\section{Quantum Spectrum of Many Anyons}

In this section we summarize the known results on the quantum mechanics of
N-anyons confined in an oscillator potential. A detailed discussion of the
results may be found in the books by Lerda\cite{lerda} and
Khare\cite{khare}. The latter contains the results for several other types
of confinement potentials as well. 

We first briefly summarize the results in the case of two anyons which is 
exactly solvable. In the case of N anyons ($N\ge 3$) the problem is not 
exactly solvable though there exists a subset of exact solutions. The 
results of this section provide the necessary framework to discuss 
further results in later sections.

\subsection{Spectrum of two-anyons}

The quantum Hamiltonian in the case of two anyons may be written as 
\begin{equation}
H = H_{cm} + H_{rel},
\end{equation}
where $H_{cm}$ is the Hamiltonian that describes the dynamics of the 
centre-of-mass. Since the statistical interaction is translation 
invariant, the centre-of-mass part is independent of this. As a result 
the spectrum of this Hamiltonian is the same as the spectrum of a two 
dimensional oscillator. That is,
\begin{equation}
E_{cm} = \hbar\omega [2n_{cm} + |l_{cm}| +1].  
\end{equation}
Here $n_{cm}$ and $l_{cm}$ denote the radial and angular momentum quantum 
numbers of the centre-of-mass excitations. 

The relative Hamiltonian is given by,
\begin{equation}
H_{rel} = \hbar\omega [ p^2 +r^2 - \alpha \frac{l}{r^2} 
+\frac{\alpha^2}{2} \frac{1}{r^2}],
\end{equation}
where $p$ is the momentum operator. 
The eigenvalue equation is easily solved by noting that the additional 
$\alpha$ dependent terms, may be combined with the centrifugal barrier 
with $l$ shifted by $l-\alpha$. The spectrum of the Hamiltonian is then 
given by,
\begin{equation}
E_{rel} = \hbar\omega [2n + |l-\alpha| +1].  
\end{equation}
Together with the spectrum of $H_{cm}$, this provides the complete 
spectrum of two anyons in an oscillator potential. Indeed this is the 
only case that is solved fully. 

We next prove the existence of a subset of exact solutions when the number
of anyons is more than two. Later we show that these solutions indeed
correspond to the quantization of collective coordinates of the N-anyon
system.  We close the discussion with an analysis of the so called
``missing states'' by specifically taking three anyons as an example. 

\subsection{Exact Solutions}

For discussing the known class of exact solutions it is convenient
to use the complex coordinates $z_i, \bar z_i$ in terms of which
the N-anyon Hamiltonian eq.(\ref{anyham}) takes the form (here after we
set $\hbar = c = 1$),
\begin{equation}
H = -2\sum_i \partial_i \bar \partial_i + \frac{1}{2} \sum_i 
z_i \bar z_i 
- \alpha \sum_{i<j}(\frac {\partial_{ij}}{\bar z_{ij}} -
\frac {\bar \partial_{ij}}{ z_{ij}}) +\frac {\alpha^2}{2}
\sum_{i \ne j,k} \frac {1}{\bar z_{ij}z_{ik}},
\end{equation}
where $\partial_i = \frac{\partial}{\partial z_i} ; ~ \partial_{ij}=\partial_i
-\partial_j$, etc.,
and the eigenvalue equation is,
\begin{equation}
H\psi(z_i,\bar z_i) = E\psi(z_i,\bar z_i).
\end{equation}
The conserved angular momentum $J$ with eigenvalues denoted by $j$, 
is given by,
\begin{equation}
J=z_i\partial_i -\bar z_i \bar \partial_i.
\end{equation}

At large distances, the oscillator confinement dominates which 
immediately signals the presence of the Gaussian factor in the wave 
function. Since the potential is 
singular as the relative distances go to zero, one needs to regulate the 
wave function in this limit to get the valid spectrum. We do this first 
by defining the variables:
\begin{equation}
X = \prod_{i<j}^N z_{ij} ; \hspace{1cm} t=\sum_{i=1}^N z_i \bar z_i,
\end{equation}
We may classify the known exact solutions as follows:

(a) $j < 0 $ solutions:
\begin{equation}
\psi_{j} = |X|^{\alpha} ~ \phi_{j}(\bar z_i) ~ e^{-t/2} ~ ;\hspace{0.5cm} j < 0
\end{equation}
with the energy eigenvalues given by,
\begin{equation}
E = N - j + \alpha \frac {N(N-1)}{2}.
\end{equation}
Here and in what follows $\phi_j$ generically denotes an eigenfunction
of the angular momentum operator $J$ with eigenvalue $j$. For
elucidating the solutions the specific form of $\phi$ is irrelevant.

(b) $j = 0$ solutions:
\begin{equation}
\psi_0 = |X|^{\alpha} ~ \phi_{0}(t) ~ e^{-t/2}  ~ ;\hspace{0.5cm} j = 0
\end{equation}
with $\phi_0(t) (=\sum_{k=1}^m C_k t^k)$ a polynomial of degree $m$ in $t$.
The corresponding energy eigenvalues are given by 
\begin{equation}
E = N + 2m + \alpha \frac {N(N-1)}{2}.
\end{equation}
The second solution is necessarily bosonic since $t$ is symmetric,
whereas the first solution
needs explicit symmetrization and antisymmetrization of the wavefunction
in terms of $\bar z_i$ to obtain the bosonic and fermionic wavefunctions.
Since this is always possible, the degeneracy of the first type of solution
is exactly the same for both bosonic and fermionic type solutions for any
given angular momentum $j$ ($ < 0 $). We may also take a combination 
(product) of the solutions of the types
discussed above, to get further $j<0$ solutions. This then generates an
infinite tower of radial excitations for each value of j.

(c) $j > 0$ solutions:
\begin{equation}
\psi_{j} = |X|^{-\alpha} ~ \phi_{j}(z_i) ~ e^{-t/2}  ~ ;\hspace{0.5cm} j > 0
\end{equation}
with the energy eigenvalues given by,
\begin{equation}
E = N + j - \alpha \frac {N(N-1)}{2}.
\end{equation}
Caution must be exercised in choosing the value of $j$  for these
solutions since the wave function is not square-integrable for all
values of $j$.  In fact the lower bound is obtained by requiring that the
wave function be square-integrable over the whole domain of $\alpha
(0\le \alpha \le 1)$. This means that $j > (N-1)(N-2)/2$. 
If however this condition is not satisfied then the wave function
remains regular only for some values of $\alpha$
($0\le \alpha \le 2j/N(N-1)$)
but not for all $0 \le \alpha \le 1$ which gives rise to the so called
non-interpolating solutions which have also been discussed in the 
literature\cite{non-int}.

All these solutions for the energy eigenvalues have a linear dependence on
$\alpha$ with a coefficient $\pm N(N-1)/2$ while the corresponding
eigenfunctions are finite order polynomials apart from the overall 
$\mid{X}\mid^{\pm \alpha}$ and the Gaussian factors. These solutions (a)--(c) cover
all the known exact solutions.  However it is by now known that these
exact solutions span only a subspace of the full Hilbert space and existence
of nonlinear solutions has been shown numerically as well as through
mean-field calculations.  It is our aim here to understand the reason for
the existence of this dichotomy.

\subsection{Missing states: Asymptotic analysis of the three-anyon
problem}

We now discuss the question of ``missing'' states, a terminology used by 
Yong-Shi Wu\cite{exact}.  The existence of these states has been shown 
numerically \cite{numeral} for three and four anyons, and through perturbative
analysis \cite{mccabe,perturb} for three anyon problem. Thomas-Fermi method has
been used to obtain approximate ground state and excitations for large
number of anyons \cite{tf}. We focus on these states now.  We cannot
easily analyze these in general for a general N-anyon system.  Therefore
we restrict ourselves to the three anyon system, where it is well known
that solutions which have nonlinear dependence on $\alpha$ exist.  Even
here, we can only comment on the nature of such solutions, but we do not
see a simple method of obtaining even low lying excitations analytically. 

Recall that the two asymptotic regions, namely $z_{ij} \rightarrow
\infty$ and $z_{ij} \rightarrow 0$ are always admissible.  Hence we
may still write the full solution as,
\begin{equation}
\psi(z_i,\bar z_i) = \exp\left(-\frac{1}{2}\sum_i z_i\bar z_i\right)
\prod_{i<j}  ~ |z_{ij}|^{\alpha} ~ \psi_2(z_i,\bar z_i),
\end{equation}
with $\psi_2$ satisfying the equation,
\begin{equation}
\left[-2\sum_i \partial_i \bar \partial_i + \sum_i 
(z_i \partial_i+ \bar z_i\bar\partial_i) 
- 2\alpha \sum_{i<j}\frac {\partial_{ij}}{\bar z_{ij}}\right]\psi_2 =
\left[E-N -\frac {\alpha^2}{2}N(N-1)\right]\psi_2
\end{equation}
The known exact solutions for the energy eigenvalues, discussed
previously, have a linear dependence on the statistical parameter $\alpha$
while the corresponding eigenfunctions are finite order polynomials, 
$\psi_2$, apart from the $\mid X \mid^\alpha$ and the Gaussian factor. 
In fact a simple
scaling argument shows that if $\psi_2$ is a polynomial
(i.e. has a finite degree), then the corresponding eigenvalue must be
linear in $\alpha$.  For, if $\psi_2$ is a polynomial in $z_i,\bar z_i$,
with highest total degree $d$ then for $z_i,\bar z_i \rightarrow \infty$ the
polynomial becomes a monomial and only the scaling operator term, 
$z_i\partial_i$, dominates. This gives,
\begin{equation}
E=N+d+\alpha \frac {N(N-1)}{2}.
\end{equation}
Thus to have non-linear dependence on $\alpha$, $\psi_2$ can not be a 
polynomial. 

Suppose $\psi_2$ admits a power series representation with infinite radius
of convergence but the series does not truncate then the scaling argument
fails.  However the analysis of such a series solution always seems to
lead to exponentially divergent behaviour making the solution
non-normalizable, ie., the power series has to truncate.  But then only
linear solutions are possible because of the scaling argument above. The
other possibility then is that the power series has a finite radius of
convergence, i.e., a ``scale'' R has to enter if nonlinearity is to be
possible. 

Let us pursue this scaling argument further, consider $\psi_2(\lambda
z_i,\lambda {\bar z_i})$ where $\lambda$ is the scaling parameter.  Having
a scale R, radius of convergence, means that $\psi_2$ has two different
representations as the scale parameter $\lambda \rightarrow 0 $ and
$\lambda \rightarrow \infty$, each being valid for $\lambda <
\lambda_{max}$ and $\lambda > \lambda_{min}$ respectively.  For $\lambda
\ge \lambda_{min}$ the scaling argument can still work but now the series
need not have only integer powers, i.e., $d$ can be a nonlinear function
of $\alpha$.  We conclude then that if a nonlinearly interpolating
eigenvalue is possible at all, the corresponding eigenfunction must have
two different series representations as $\lambda \rightarrow 0$ and
$\lambda \rightarrow \infty$.  One then has to try to match the two series
suitably. 

Exploring this possibility is extremely complicated for general N.  The
first non-trivial case of $N = 3$ is analyzed in \cite{dkm}. While this
analysis shows that the possibility of finite radius of convergence is
viable, one is not able to obtain any specific non-linearly interpolating
solution.  Observe that to get a discrete spectrum, the solutions of the
eigenvalue equation must be regular and normalizable. For a single power
series, this forces us to truncate the series and one obtains linearly
interpolating spectrum. For multiple power series for the eigenfunctions,
while non-linear dependence is conceivable, it is hard to see
regularity/normalizability to get discrete spectrum. One can get a glimpse
of the intricacies through the following analysis. 

Consider the method of analyzing solutions in the three anyon problem 
consists of using the Fourier expansion of the wavefunction.
Since the center of mass degree of freedom is irrelevant for further 
arguments, we may separate that out and write the wavefunction as
$$ \psi = \exp\left(-\frac{1}{2}\sum_i r_i^2\right)
~ |X|^{\alpha}~ \phi(r_1, r_2, \theta_1, \theta_2) ~ \Phi_{CM}(R,\Theta),$$
where $\vec r_1, \vec r_2$ are relative coordinates chosen appropriately. 
In general $\phi$ may be written as,
\begin{equation}
\phi_j(r_1,r_2,\theta_1,\theta_2)=
\sum_{n_1,n_2} e^{i n_1\theta_1}e^{i n_2\theta_2}~ \chi_{n_1,n_2}(r_1,r_2)
=\sum_{n} e^{i j \frac{(\theta_1+\theta_2)}{2}}
e^{i n \frac{(\theta_1-\theta_2)}{2}}~ \chi_{j,n}(r_1,r_2)
\end{equation}
for a given angular momentum $j$, where $j=n_1+n_2$, $n=n_1-n_2$.  We can 
now use this
representation to look at the solutions for a given $j$. The eigenvalue
equation now becomes $H_{rel}\phi_j = {\cal E}\phi_j $, where $H_{rel}$
is the three-anyon Hamiltonian in relative coordinates and ${\cal E} =
E-2-3\alpha$ as before.  Substituting for $\phi_j$ the eigenvalue equation
takes the form,
\begin{eqnarray*}
& &3x_2\left[4x_1\frac{\partial^2}{\partial x_1^2}+4(1+\alpha-x_1)\frac{\partial}
{\partial x_1} -\frac{(j+n)(j+n-4\alpha)}{4x_1}\right.\\
&+&\left.4x_2\frac{\partial^2}{\partial x_2^2}+4(1+2\alpha-x_2)\frac{\partial}
{\partial x_2} -\frac{(j-n)(j-n-8\alpha)}{4x_2}+2{\cal E}\right]\chi_{j,n}\\
&=&x_1\left[4x_1\frac{\partial^2}{\partial x_1^2}+4(1+3\alpha-x_1)\frac{\partial}
{\partial x_1} -\frac{(j+n+4)(j+n+4-12\alpha)}{4x_1}\right.\\
&+&\left.4x_2\frac{\partial^2}{\partial x_2^2}+4(1-x_2)\frac{\partial}
{\partial x_2} -\frac{(j-n-4)^2}{4x_2}+2{\cal E}\right]\chi_{j,n+4},
\end{eqnarray*}
where $x_i=r_i^2$.  This expression is exact, representing an infinite set of 
coupled equations. They relate Fourier modes differing by 4, i.e., if 
$\chi_0,\chi_1,\chi_2,\chi_3$ are known then $\chi_{4k},\chi_{4k+1},
\chi_{4k+2}, \chi_{4k+3}$ get determined in terms of $ \chi_0,...,\chi_3$
etc.  However there is no relation among $\chi_0,\chi_1,\chi_2,\chi_3$
themselves.  In a sense these four functions will give four independent
solutions of the eigenvalue equations.  We can then deal with a given 
``tower'' separately and independently and this is true
for every given $j$.  Let us concentrate on one tower. Now three distinct
cases arise naturally:
\begin{flushleft}
(a) Only one member of the tower is nonzero, ie., $\chi_{j,n} = \chi_{j,m}
\delta_{n,m}$.

(b) Only a finite number of $\chi_{j,n}$'s are nonzero, ie., $\chi_{j,n}
=0~~ \forall ~~n \ge n_1$ and $\forall ~~  n \le n_2$ with $ n_2 < n_1$.

(c) Infinitely many $\chi_{j,n}$'s are nonzero.  This gives raise to the
following cases:
\begin{eqnarray*}
\chi_{j,n}&=&0 ~~\forall~~ n\ge n_1 \\
\chi_{j,n}&=&0 ~~\forall~~ n\le n_2 \\
\chi_{j,n}&\ne& 0 ~~\forall~~ n.
\end{eqnarray*}
\end{flushleft}

Case (a) is simple to analyze and it reproduces the known exact solutions. 
In the exact eigenvalue equation relating $\chi_{j,n}$ and $\chi_{j,n+4}$
given above, we set the left hand side to zero which gives an equation for
the $\chi_{j,n}$.  But the right hand side also gives rise to another
equation for the same $\chi_{j,n}$ when the coupling to $\chi_{j,n-4}$ is
taken into account.  However the form of the equations suggests that the
two equations are separable but must be consistently solved.  The
consistency conditions immediately yield the linear exact solutions which
are already outlined above.  Cases (b) and (c) seem quite complicated and
nonlinearly interpolating states must be in one of these cases. Consider
the normalization condition on the full wavefunction, 
\begin{equation}
\parallel \psi_j \parallel^2 = \sum_n \int_0^{\infty} dx_1 dx_2
|X|^{2\alpha} |\chi_{j,n}|^2 e^{-{x_1+x_2}} \equiv \sum_n C_n.
\end{equation} 
If $\psi_j$ is given by case (b) then there are only a
finite number of terms in the norm and if $C_n$'s are finite then $\psi_j$
is normalizable.  Thus the quantization condition for the nonlinear states
must arise by the demand that $C_n$ must be finite.  For the case (c) even
if all $C_n$'s are finite we may still get the norm of the wavefunction to
be infinite because the sum above may not converge and then the
quantization condition would be the convergence of $\sum_n C_n$. It is
therefore apparent that non-linear dependence on $\alpha$ must come about
in a very complicated manner. 

\section{Classical Analysis of Many Anyons and Integrability}

Having noted the difficulties inherent in describing the non-linearly
interpolating solutions, it is striking that the subspace spanned by the
exact solutions are obtained so easily. Obviously these two sets must have
very different origins. Indeed as we will see below, the exact
solutions have their origin in the quantization of collective degrees of
freedom. To explore this aspect we analyze the system classically.

We may now, therefore,  ask the question, what is the classical Lagrangian 
which gives raise to the Hamiltonian of Eq.(6)?  In fact it is quite
straight forward to see that the Lagrangian is given by,
\begin{equation}
L=\frac{1}{2}\sum_{i=1}^{N} ~  (\dot{\vec{r}} \,_i)^2 + \alpha \sum_{i<j}^{N}
\dot{\theta}_{ij} -V(\vec r_i);\hspace{1cm}
\theta_{ij}=tan^{-1}\left(\frac{y_i-y_j}{x_i-x_j}\right),
\label{lagrange}
\end{equation}
where $\vec r_i$ denote particle coordinates and $V(\vec r_i)$ is some
confining potential which we choose to be the harmonic oscillator
potential $V(\vec r_i)= \sum_{i=1}^N \vec r_i \,^2$.  The $\alpha$-dependent
term is the statistical interaction.   The harmonic oscillator
potential is convenient since the dynamics of the system is well
understood in the limit $\alpha = 0$. As mentioned in the
previous section, the corresponding quantum system displays the existence 
of a subset of exact solutions, though the full spectrum is not
analytically solvable for many anyons.
It is possible that the system is at least partially separable to allow
for the known exact solutions.  We will return to this question to the
next sub-section and consider the question of integrability first.  If one
considers the classical Lagrangian, the statistical interaction is a
total time derivative.  The Euler-Lagrange equations of motion are
therefore the same as those of a 2N-dimensional oscillator and this
of course is an integrable system.  However the quantum spectrum
displays not only the exact solutions but it has been demonstrated
numerically in the case of three and four anyons that the spectrum
also shows evidence of many level crossings and level
repulsions \cite{numeral,djm} which is often taken to be an indication of
non-integrability. Here we  argue that even though we have $2N$ constants 
of motion in involution the invariant surfaces do not have the topology of 
a $2N$ dimensional torus. 

We begin with the analysis of the classical Lagrangian given by
Eq.(\ref{lagrange}). It is convenient to write the Lagrangian in the form,
\begin{equation}
L=\frac{1}{2}\sum_{i=1}^{N} \left[(\dot{\vec r}_i)^2 - \vec r_i\,^2\right]
+ \alpha \sum_{i<j}^{N} \frac {\vec r_{ij} \times \dot{\vec {r}}_{ij}}
{\vec r_{ij} \,. \,\vec r_{ij}},
\end{equation}
where the dots indicate the time derivatives.
The first step is to introduce relative coordinates, also called Jacobi 
coordinates and
separate the trivial centre of mass degree of freedom. To this end we
write,
\begin{equation}
\vec{\rho}_a=\frac{1}{\sqrt{a(a+1)}}\sum_{k=1}^{a}
\vec{r}_k-\sqrt{\frac{a}{a+1}}\vec{r}_{a+1};~~~a=1,...,N-1 ;
\end{equation}
with the inverse relation given by,
\begin{equation}
\vec{r}_i=\left[-\sqrt{\frac{i-1}{i}}\vec{\rho}_{i-1}
+\sum_{k=i}^{N-1}\frac{\vec{\rho}_k}{\sqrt{k(k+1)}}\right] +\vec{R}_{cm} \equiv
A_{i}^{a}\vec {\rho}_{a}+\vec{R}_{cm},  
\end{equation}
where $\vec{\rho}_{a},\, a = 1,...,N-1$ are dimensionless relative coordinates and
$\vec{R}_{cm}$ is the centre of mass coordinate.
It follows that,
$$ \sum_i A_i^a =0; \hspace{1cm} \sum_i A_i^a A_i^b = \delta^{ab}.$$
It is straight forward to see that,
$$ L=L_{CM}+L_{rel},$$
where 
$$ L_{CM}=\frac{1}{2}[\dot {\vec R}_{CM}^2 -\vec R_{CM}^2],$$
$$ L_{rel} = \frac{1}{2} \left[ \dot {\vec \rho}_a\,^2 - \vec \rho_a^2 \right] 
+ \alpha \sum_{i<j} \frac{A_{ij}^a \, A_{ij}^b ~ \vec \rho_a \times \dot{ \vec
\rho}_b}{A_{ij}^c \, A_{ij}^d ~ \vec \rho_c . \vec \rho_d},$$
where $A_{ij}^a = A_i^a -A_j^b$.  From now on we concentrate only on
the $L_{rel}$ and drop the subscript.  It is easy to see that the 
Euler-Lagrange equations of motion are,
\begin{equation}
\ddot{\vec \rho}_a = -\vec \rho_a.
\end{equation}
There is no $\alpha$ dependence in these equations since the
corresponding term is a total derivative.  From these equations it
follows that
\begin{equation}
{\cal E}_a \equiv \frac{1}{2} [\dot{\vec{\rho}}_a^2+\vec{\rho} _a^2];
\hspace{0.5cm} l_a \equiv \vec{\rho} _a \times \dot{\vec{\rho}}_a; \hspace{0.5cm}
a=1,...,N-1
\end{equation}
are constants of motion.  In the Hamiltonian formulation the $\dot
{\vec{\rho}}_a$ are expressed in terms of conjugate momenta and coordinates
which {\it do contain} the $\alpha$ dependence,
\begin{eqnarray}
P_{ax}=\frac{\partial L}{\partial \dot{\rho}_{ax}}=\dot{\rho}_{ax}
-\alpha {\cal A}_{ax}; \hspace{0.5cm}
{\cal A}_{ax} = \sum_{i<j} \frac{A_{ij}^a \, A_{ij}^b ~ \rho_{by}}
{A_{ij}^c \, A_{ij}^d ~ \vec \rho_c . \vec \rho_d},\\
P_{ay}=\frac{\partial L}{\partial \dot{\rho}_{ay}}=\dot{\rho}_{ay}
+\alpha {\cal A}_{ay}; \hspace{0.5cm}
{\cal A}_{ay}= \sum_{i<j} \frac{A_{ij}^a \, A_{ij}^b ~ \rho_{bx}}
{A_{ij}^c \, A_{ij}^d ~ \vec \rho_c . \vec \rho_d},
\end{eqnarray}
and the constants of motion in relative coordinates are,
\begin{eqnarray}
{\cal E}_a & = & \frac{1}{2}\left[(P_{ax}+\alpha {\cal A}_{ax})^2
+(P_{ay}-\alpha {\cal A}_{ay})^2 +\vec {\rho}_a\,^2\right] \nonumber \\
l_a & = & \vec{\rho}_a\times \vec{P}_a -\alpha (\rho_{ax}{\cal A}_{ay}
+\rho_{ay}{\cal A}_{ax}).
\end{eqnarray}
Clearly the Hamiltonian is given by
$$ H = \sum_{a=1}^{N-1} {\cal E}_a.$$
Note that the $\alpha$-dependent term in $L$ is well defined only if
$\vec{r}_{ij}\ne 0$ for all $i,j$,  that is, only if 
$$A_{ij}^a \, A_{ij}^b ~ \vec{\rho}_a.\vec{\rho}_b ~\ne~ 0 ~~\forall~~ i,j.$$  
Consequently the expressions
for ${\cal E}_a, l_a$ and $H$ are also valid only if $\vec{r}_{ij}$ is
not zero.  In effect the classical configuration space on which $L$ is
well defined is the space 
\begin{equation}
Q^{N-1} = R_2^{N-1}-\Delta; \hspace{0.5cm} \Delta = \{~\vec{\rho}_a~/
~A_{ij}^c A_{ij}^d~ \vec{\rho}_c.\vec{\rho}_d~ =~0 {\rm ~~
for~ some~ pair(s)}~~i,j\}.
\end{equation}
The space $Q^{N-1}$ is not simply connected.  Its fundamental 
group $\pi_1$ is the same as the fundamental group of $\{R_2 - (N-1)\}$
points which is known to be nontrivial.  For $N=2$, $\pi _1 =
Z$ while for $N \ge 3,~~ \pi_1$ is non-Abelian.

The corresponding phase space is the cotangent bundle of $Q$ on which 
${\cal E}_a, l_a$ and $H$ are well defined.  This phase space is
topologically nontrivial and for our purposes we do not need the full
machinery for handling this topologically nontrivial phase space. It
is sufficient to pretend that the full space is still $R^{N-1}_2$ but
simply avoid coincident points.

It is straightforward then to prove the following Poisson bracket
relations, 
\begin{eqnarray}
\{{\cal E}_a,{\cal E}_b\} & = &\{l_a,l_b\} = \{{\cal E}_a,l_a\}=0
\hspace{0.3cm} \forall \hspace{0.3cm} a,b=1,...,N-1, \\
\{H,{\cal E}_a\}&  = & \{H,l_a\} =  0.
\end{eqnarray}
Thus for $4(N-1)$ degrees of freedom $P_a, \rho_a$ we have 
$2(N-1)$ constants of motion in involution.  So a necessary condition
for a system to be integrable is satisfied.  We will tentatively refer
to this system as being integrable (or potentially integrable) in the
``Liouville sense''. 
However, for the system to be integrable via
action angle variables, further conditions have to be
satisfied (see page 271 of \cite{arnold}):  If
$M({\cal E }_a,l_a)$ denote the set of points in phase space at which
the constants of motion have values ${\cal E }_a,l_a$ then (a) $M$ is
a $2(N-1)$ dimensional submanifold iff ${\cal E }_a,l_a$ are all
independent, (b) if $M$ is a submanifold which is compact and
connected  then $M$ is
a $2(N-1)$ dimensional torus. If these conditions are satisfied then
one can introduce action-angle variables in the neighbourhood of $M$.
The orbits on $M$ will all be conditionally periodic in general and
for the periodic orbits  one may use Bohr-Sommerfeld
quantization to get a subset of eigenvalues.

We will show in the next section that there exist $M({\cal E
}_a,l_a)$ for which action-angle variables can be introduced.  These
correspond to the collective motion of N-anyons and their
semiclassical quantization reproduces the exact eigenvalues of the quantum
system that are already known. However,
if we go away from these initial conditions some of the orbits of
the
oscillator will cease to be periodic when $\alpha$ dependence is
introduced.  For instance the Euler-Lagrange equations 
$$\ddot{\vec{r}}_i=-\vec{r}_i \Rightarrow
\ddot{\vec{r}}_{ij}=-\vec{r}_{ij} ~~({\rm independent~ of~} \alpha).$$
For any pair $ij,~i\ne j$, $\vec{r}_{ij}$ in general describes an
ellipse in the configuration space.  However for zero orbital angular
momentum, the ellipse degenerates to a straight line, hence the point
$\vec{r}_{ij}=0$ is on such an orbit.  
Since these points are not admissible when $\alpha
\ne 0$ these orbits cannot lead to periodic orbits in the phase space.
For all choices of ${\cal E}_a, l_a$ such that for some initial
conditions the phase space orbits will have $r_{ij}$ approaching
arbitrarily close to zero, $M({\cal E}_a,l_a)$ cannot be a torus.
(Even if $M$ is a manifold it will fail to be compact and/or connected.) 
In appendix A, we consider the explicit example of an angular momentum
zero orbit in the case of two anyons, which necessarily passes through
the diagonal  point.  Even though in configuration space this may
appear to be
periodic, in the phase-space the orbits stay on the surface of a
cylinder which can be broken into two disconnected sheets.  
Therefore for
$\alpha\ne 0$, although we have a potentially integrable system it is
not generically integrable via action-angle variables.
The existence of such an 
$M$ indicates the possibility of  ``extreme sensitivity'' to the initial
conditions.  This is a possible reason for the ``level repulsions'' seen
numerically. This bears a resemblance to the billiard system
considered by Richens and Berry \cite{berry} though the reasons 
for the failure of integrability via
action-angle coordinates are different.
Following Richens and Berry we conclude that our system
is {\it Pseudo-integrable}. We will sharpen and formalize this notion later in 
Section 5.

\subsection{Collective modes and relation to exact states}

We may now ask the question, what causes these states, though a subset of
the full Hilbert space, to be exact solutions of the N-anyon Hamiltonian
which is in general non-separable. Indeed as we will show now, these
solutions may be obtained by quantizing the collective coordinates of the
N-anyon system in relative coordinates consistent with the equations of
motions for a set of initial conditions.  In some sense these collective
coordinates do not give us any information on the internal dynamics of the
system.  These are therefore trivial solutions dynamically.  The anyonic
interaction term in $L$ is invariant under two sets of time independent
transformations: 
\begin{eqnarray}
\vec{\rho}_a &\rightarrow & \vec{\eta}_a = U(\theta)\vec{\rho}_a
\hspace{0.3cm} \forall \hspace{0.3cm} a,\\
\vec{\rho}_a &\rightarrow & \vec{\eta}_a = \lambda\vec{\rho}_a
\hspace{0.3cm} \forall \hspace{0.3cm} a,
\end{eqnarray}
where $U(\theta)$ denotes a rotation of the vector by an angle
$\theta$ in two dimensions.
Therefore for any given configuration $\{\vec{\rho}_a\}$ at any given
time t we can always rotate the axes so that  $\rho_{1y}=0$ (say).  Thus
by making a time dependent rotation we can ensure that $\eta_{1y} =0$
for all $t$.  Defining, 
\begin{equation}
\vec{\rho}_a =
\begin{pmatrix} 
 \cos\theta & -\sin\theta \\ \sin\theta & \cos\theta 
\end{pmatrix}
\vec{\eta}_a,
\end{equation}
The following identities are easy to prove:
\begin{eqnarray}
\vec {\rho}_a.\vec {\rho}_b & = & \vec {\eta}_a.\vec {\eta}_b ,\\
\vec {\rho}_a \times \vec {\rho}_b & = &
\vec {\eta}_a \times \vec {\eta}_b ,\\
\dot{\vec {\rho}}_a\,^2 & =& \dot{\vec {\eta}}_a\,^2 + 2\dot\theta \,\vec {\eta}_a
\times \dot{\vec {\eta}}_a+\dot{\theta}^2\,\vec {\eta}_a\,^2,\\
\vec {\rho}_a \times \dot{\vec {\rho}}_b & = &
\dot{\theta}\,\vec {\eta}_a.\vec {\eta}_b + \vec {\eta}_a \times
\dot{\vec {\eta}}_b.
\end{eqnarray}
Now the set $\{\eta_a ; a=1,...,N-1\}$ is effectively a $2(N-1)-1$
dimensional vector since $\eta_{1,y}=0$ for all time t.  We can
therefore introduce the standard spherical coordinates by the usual
procedure,
\begin{equation}
\vec{\eta}_a \equiv R \vec{\xi}_a; \hspace{0.3cm}\xi_{1y}=0;
\hspace{0.3cm} \sum_a \xi_a^2 = 1,
\end{equation}
where
\begin{eqnarray}
\xi_{N-a,x} & = &s_1 s_2 ...s_{2a-2}c_{2a-1},\\
\xi_{N-a,y} & = &s_1 s_2 ...s_{2a-1}c_{2a},\\
\xi_{1,x} & = &s_1 s_2 ...s_{2N-4},\\
\xi_{1,y} & = & 0,
\end{eqnarray}
where $a=2,...,N-1$ and $s_{\mu}=\sin \theta_{\mu},~
c_{\mu}=\cos \theta_{\mu}$.
In terms of these variables the Lagrangian may be
rewritten as 
\begin{eqnarray}
 L&=&\frac{1}{2}\left[\dot {R}^2 - R^2 + R^2\dot{\theta}^2 + 2R^2
\dot{\theta}\sum_a\vec{\xi}_a \times \dot{\vec {\xi}}_a + R^2 \sum_a
\dot {\vec{\xi}}_a\,^2 \right] \nonumber\\
 &+& \alpha \frac{N(N-1)}{2}\dot {\theta}+
\alpha \sum_{i<j} \frac{A_{ij}^a \,A_{ij}^b \vec \xi_a \times ~ \dot{\vec
\xi}_b}{A_{ij}^c \, A_{ij}^d \vec \xi_c . \vec \xi_d}.
\end{eqnarray}

Since $\vec\eta_a$ are obtained from $\vec\rho_a$ by the same rotation
matrix for all $a$, the angle $\theta(t)$ clearly describes a collective
rotation of all the N anyons about some reference axis.  The anyonic
interaction term ( a total time derivative) is manifestly independent
of $R(t)$. This may also be regarded as a collective mode as discussed
below.  It is the semiclassical quantization of these two modes that
yields the exactly known energy eigenvalues.   To elaborate these
points let us consider the Euler-Lagrange equations of motion.  Since
the $\alpha$ dependent part of the Lagrangian is a total time
derivative we can ignore it for analyzing the classical equations of
motion.  Clearly these equations are identical to that of the
oscillator equations of motions.  Translating this
into the spherical coordinates we obtain,
\begin{equation}
[\ddot{R}+R-R\dot{\theta}^2]\vec{\xi}_a-[R\ddot{\theta}+2\dot{R}
\dot{\theta}]V\vec{\xi}_a+2[\dot{R}-R\dot{\theta}V]\dot{\vec{\xi}}_a
+R\ddot{\vec{\xi}}_a =0,
\label{eq50}
\end{equation}
where
$$V=\begin{pmatrix}
     0 & 1 \\ -1 & 0
    \end{pmatrix}$$
and in the matrix equation above the $\vec{\xi}$ is a column vector
with two elements.  The matrix V between two vectors essentially
generates their cross product.  Taking the dot product with
$\vec\xi_a$ and summing over $a$ we find,
\begin{equation}
[\ddot{R}+R-R\dot{\theta}^2]=2R\dot{\theta}\sum_a\vec{\xi}_a \times
\dot{\vec{\xi}}_a +R\sum_a\dot{\vec{\xi}}_a^2 ,
\end{equation}
where we have made use of the identities,
$$ \sum_a \vec{\xi}_a.\dot{\vec{\xi}}_a=0, \hspace{0.3cm} \sum_a
\vec{\xi}_a.\ddot{\vec{\xi}}_a=-\sum_a \dot{\vec{\xi}}_a^2.$$
Taking the cross product with
$\vec\xi_a$ and summing over $a$ we find,
\begin{equation}
[R\ddot{\theta}+2\dot{R}\dot{\theta}]=-2\dot{R}\sum_a
\vec{\xi}_a \times
\dot{\vec{\xi}}_a -R\sum_a\vec{\xi}_a \times
\ddot{\vec{\xi}}_a.
\end{equation}
Using the above two relations Eq.(\ref{eq50}) can be rewritten as,
\begin{equation}
R\ddot{\vec{\xi}}_a
+2[\dot{R}-R\dot{\theta}V]\dot{\vec{\xi}}_a
+[2R\dot{\theta}\vec{\xi}_b \times
\dot{\vec{\xi}}_b +R\dot{\vec{\xi}}_b^2]\vec{\xi}_a+
[2\dot{R}\vec{\xi}_b \times
\dot{\vec{\xi}}_b +R\vec{\xi}_b \times
\ddot{\vec{\xi}}_b]V\vec{\xi}_a
=0,
\end{equation}
where sum over $b$ is assumed.
Here the equations of motion for $R$ and $\theta$ are still coupled to
all the other internal coordinates, $\vec \xi_a$ and as yet they are not collective
coordinates.  We therefore  need to impose a set of initial conditions
which
may lead to the separation of these two modes as collective modes from
the internal variables $\theta_{\mu}$.  To realize this let at time
$t=0$, all velocities $\dot{\vec{\xi}}_a(t=0) =0$ for all $a$.  Then the
above equations reduce to, 
\begin{eqnarray}
R[\ddot{\vec{\xi}}_a +(\sum_b \vec{\xi}_b \times
\ddot{\vec{\xi}}_b) V\vec{\xi}_a]=0, \\
\ddot{R}+R-R\dot{\theta}^2=0, \\
R\ddot{\theta}+2\dot{R}\dot{\theta}=-R\sum_b \vec{\xi}_b \times
\ddot{\vec{\xi}}_b.
\end{eqnarray}

Now consider the first equation for $R\ne 0$ and $a=1$.  From the initial
conditions it is obvious that,
$$\ddot{\xi}_{1x}=0; \hspace{0.5cm}\xi_{1x} 
\sum_b (\vec{\xi}_b\times\ddot{\vec{\xi}}_b) = 0$$
because $\xi_{1y}=0 ~~\forall~~ t$. 

Now {\it if} $\xi_{1x}$ is nonzero, then 
$\sum_b\vec{\xi}_b \times\ddot{\vec{\xi}}_b=0,$ and hence 
$$\ddot{\xi}_b=0~~\forall~~b.$$
Since the equations of motion are second order in time t, we have proved that:
{\it if} $\dot{\vec{\xi}}_a(0) = 0 ~~\forall~~ a ,~~ and~~ R(0) ,~
\xi_{1x}(0) $ are
non zero {\it then} $ \forall~~ t $,
$$\vec{\xi_a}(t) = \vec{\xi_a}(0) , $$
$$\ddot{R} + R - R {\dot{\theta}^2} = 0, $$
$$R {\ddot{\theta}} + 2 {\dot{R}} {\dot{\theta}} = 0.$$
For oscillator ($\alpha = 0$) R(0) and $\xi_{1x}(0) $ being non zero is
a special class of initial condition. 
Indeed in general for a second order equation system if both first and
second order derivatives vanish at any $t$ then this can hold only for
some restricted class of coordinate values. For the equations of motions
considered in the configuration space $R^{N-1}_2$ the restricted class
is precisely characterized by $ R(0) \ne 0 $ and $ \xi_{1x}(0) \ne 0 $.
However, when $\alpha \ne 0$ the
configurations space is $Q^{N-1}$ which does not contain points which
have R = 0 and $\xi_{1x} = 0 $. Thus there are no restrictions on the
initial conditions in $Q^{N-1}$.

The above initial conditions amount to freezing the ``internal motion'' of
the anyons. The remaining motion is a collective motion described by
R(t) and $\theta(t)$ which is just a $2(N-1)$ dimensional oscillator. R(t)
being nonzero implies that the angular momentum must be nonzero. This
explains in what sense R(t) can be interpreted as a collective mode.
For describing the motion of collective mode the effective Lagrangian
is 
\begin{equation}
L_{eff}= \frac{1}{2} [\dot{R}^2 + R^2 \dot{\theta}^2 -R^2] +\alpha
\frac{N(N-1)}{2} \dot{\theta}.
\end{equation}
This is identical to the Lagrangian in  the relative coordinate for a
two anyon system with $\alpha \rightarrow \alpha \frac{N(N-1)}{2}$
and in $2(N-1)$ dimensions.
Semiclassical quantization will then reproduce all exactly known
energy eigenvalues summarized in Sect. 3.  
In this effective Lagrangian picture the only
known memory of N resides in the coefficient of $\alpha$.  This can be
understood by noting that when all the N-anyons are rotated by $2\pi$
about the centre of mass they also circle each other to pick up the
extra phase.  

If on the other hand  we consider 
$\dot{R}= \dot{\theta}=0$ at $t=0$ class of initial
conditions then we see that $\ddot{R}, \ddot{\theta}$ are nonzero and
hence $R(t), \theta(t)$ do depend on $\theta_{\mu}$.  Thus the
collective motion is not fully decoupled from the ``internal motion''.
We therefore refer to this system as partially separable.

We also note that the same conclusions may be drawn from the Hamiltonian
formulation. For completeness we give the relevant expressions and
arguments.  The conjugate momenta for the hyper-spherical variables
are given by, 
\begin{eqnarray}
P_R & = & \dot{R},\\
P_{\theta}& = & R^2 \dot{\theta} + R^2 \sum_{\mu}
\dot{\theta}_{\mu}F_{\mu} +\alpha
\frac{N(N-1)}{2}, \\
P_{\mu} & = & R^2 r_{\mu}^2 \dot{\theta}_{\mu} + 
R^2 \dot{\theta}F_{\mu} +\alpha G_{\mu}, 
\end{eqnarray}
where,
$$ r_{\mu}=s_1 s_2...s_{\mu-1},~~\mu=1,...,2N-3,~~r_1=1$$
and $F_{\mu}$ and $G_{\mu}$ are defined through,
$$ \sum_{\mu =1}^{2N-4}\dot{\theta}_{\mu}F_{\mu} \equiv
\sum_{a=1}^{N-1}\vec{\xi}_a  \times \dot{\vec {\xi}}_a$$
$$ \sum_{\mu =1}^{2N-4}\dot{\theta}_{\mu}G_{\mu} \equiv
\sum_{i<j} \frac{A_{ij}^a \, A_{ij}^b ~ \vec \xi_a \times \dot{\vec
\xi}_b}{A_{ij}^c \, A_{ij}^d ~ \vec \xi_c . \vec \xi_d}.$$
The Hamiltonian is then given by 
\begin{equation}
H=H_1+H_2,
\end{equation}
where
\begin{eqnarray}
H_1 & = & \frac{1}{2} [P_R^2 + R^2] + \frac{(P_{\theta}-\alpha
\frac{N(N-1)}{2})^2}{2R^2}, \\
H_2 &=& \frac{1}{2R^2} \left[ \sum_{\mu} \frac{(P_{\mu}-\alpha
G_{\mu})^2}{r_{\mu}^2} +\frac{(P_{\theta}-\alpha
\frac{N(N-1)}{2}-\sum_{\mu}\frac{F_{\mu}(P_{\mu}-\alpha
G_{\mu})}{r^2_{\mu} 
})^2}{1-\sum_{\mu}\frac{F_{\mu}^2}{r_{\mu}^2}} \right. \nonumber\\ 
&& \left. ~~~~~~~~ - \, (P_{\theta}-\alpha
\frac{N(N-1)}{2})^2 \right]. 
\end{eqnarray}
It is easy to see that for the special initial conditions
$\dot{\theta}_{\mu} = 0 $, $H_2 =0$.  Also the Poisson bracket
$\{H_1,H_2\} \propto H_2$
vanishes for these initial conditions.
The corresponding quantum statement would $[H_1,H_2] \propto H_2$,
and therefore  
if we consider the
subspace of states $\{\psi\}$ on which $H_2\psi =0$ 
then $H_1$ will act invariantly on this subspace.
Further $H=H_1$ on this subspace and thus the 
eigenstates of $H_1$ will be exact eigenstates of the full
system and conversely. But the
problem of solving $H_1$ is analogous to that of two anyon problem
with $\alpha \rightarrow \alpha \frac{N(N-1)}{2}$.  
Therefore these solutions
are given by,
\begin{eqnarray}
E & = & 2m + |j - \alpha \frac{N(N-1)}{2}| + (N-1), \\
\psi & = & C \, R^{|j-\alpha|} \, e^{-R^2/2} ~ f(R),\\
H_1 \psi & = & E \psi
\end{eqnarray}
where $f(R)$ is some polynomial of degree $2m$ in $R$.  In general the
normalization constant C in
$\psi$ will have dependence on $\theta_{\mu}$'s, restricted by
$H_2\psi=0$.   These are
precisely all the known solutions and the above argument shows that
there are no more solutions  satisfying $H_1 \psi = E \psi $ and $H_2
\psi = 0$.  Thus $H_2 \psi = 0 $ characterizes the subspace spanned by
{\it all} the known exact solutions.  So the quantum counterpart of
classical partial separability implies the existence of these
solutions.   

\section{Semiclassical Analysis --- Periodic Orbit Theory}

We have seen the difficulty in obtaining the non-linearly interpolating
eigenvalues. We have also discussed their possible relation to, {\it {at
classical level}}, pseudo-integrability on the one hand and the
possibility of two different asymptotic series at the quantum level on the
other hand. There is yet another possible approach to gain some
understanding of these eigenvalues, namely semi-classical analysis. This
is what we discuss now. We begin by classifying the classical orbits for
N-anyon system and discuss the trace formula in the subsequent section.
Most of the discussion of this section is based on reference \cite{gdd}.

\subsection{Classification of orbits and symmetry reduction}

We will first carry out a classification of orbits for $N$
anyons and see how the pseudo-integrability amounts to a reduction of the
dynamical symmetry. We will also see the role played by the fundamental
group of the phase space in this regard. Noting that a trajectory is
specified by giving an initial point, we will use the dynamical symmetry
of the oscillator system to group various orbits into continuous families.
For anyons, certain points and hence certain orbits are disallowed
(removed) with the result that one gets several continuous families. 

\subsubsection{The case of two anyons}

As usual we can separate the center-of-mass (CM) dynamics from the relative
coordinate dynamics. The CM dynamics is trivial -- identical to an
oscillator locally and globally. The anyonic feature is contained entirely in
the relative dynamics described as:
\begin{equation}
\begin{array}{lclclcl}
Q & = & R^2 - \vec{0} &~~~ ,~~~ & \Gamma & = & Q \times R^2 \\
L & = & \frac{1}{2} ( \dot{\vec{r}}^2 - \vec{r}^2 ) + \alpha_{c} \dot{\theta} &
~~~,~~~ & \theta & \equiv & \tan^{-1}{y/x} \\
H & = & \frac{1}{2}(\vec{\pi}^2 + \vec{r}^2) &~~~ ,~~~ & \vec{\pi} & \equiv & 
\vec{p} - \alpha_c \frac{\hat{k} \times \vec{r}}{r^2}  
\end{array}
\end{equation}

The classical trajectories are trivially known in these cases. These are
ellipses with two constants of motion, the energy ($E$) and the angular 
momentum ($J \ = \ (\vec{r} \times \dot{\vec{r}})_z $). The sign of $J$ gives
the sense of traversal and a trajectory reaches the point $\vec{r} \ = \ \vec{0}$
if and only if $J \ = \ 0$. These degenerate trajectories ($J \ = \ 0$) are
not allowed for the anyons though are allowed for the oscillator. 

Consider the set of all possible trajectories with a given energy. Fix two 
such trajectories with angular momenta $J, J^{\prime}$. Suppose there exists a
continuous family of interpolating trajectories connecting these two. Clearly
if the angular momenta have opposite signs, then at least one of the
interpolating trajectories must be degenerate. But this is disallowed for
anyons. Hence, for anyons there must be {\it {at least}} {\bf two} `orientation' classes 
of trajectories distinguished only by the sign of angular momenta.

While we have shown that there must be at least two families we do not know 
yet if there {\it {must}} be {\it {precisely}} two families. We will use the 
symmetries to show that there are precisely two families for anyons and 
precisely one family for the oscillator. 

As far as the PB algebra is concerned, the oscillator and anyons
are identical and hence we do have {\it {infinitesimal}} symmetries forming the
Lie algebra of $U(2)$ (or $OSp(4,R)$) for both cases. The constant $H$
surface ($S^3$ in the phase space $R^4$ for oscillator) is of course
invariant under the infinitesimal symmetry transformations. For anyons, the
phase space has $\vec{r} = \vec{0}$ removed and the constant $H$ surface has
a great circle (one dimensional) removed from the $S^3$, say a ``longitude". 

In the appendix B we have given the details of the $OSp(4N,R)$ symmetry of the
oscillator. In the present case of two anyons with center of mass coordinate
separated, we have effectively, $N \ = \ 1$ and the symmetry generators form
$OSp(4, R)$. We have only the 4 generators of the ${\bf {T}}_i$ type.
Of these four, ${\bf u}_1$ generates rotations of position and
momentum while ${\bf u}_2$ generates time evolution. The generic integral 
curves are given in the appendix B. 

From appendix B we also recall that $OSp(4, R)$ acts transitively on $H = E$ 
sphere.
This shows that for the oscillator there is precisely ONE family of
trajectories. For anyons we need analogous result with the extra condition
that degenerate trajectories are avoided to conclude that there are precisely 
{\bf two} families. 

{\underline {Claim}}: Given any two trajectories with the same sign of their
angular momenta, there exists a one parameter subgroup connecting the two 
without changing the sign of angular momenta. 

{\underline {Proof}}: Observe that ${\bf u}_1, {\bf u}_2$ transformations
leave the angular momentum invariant. Choose ${\bf u} \ = \ \cos(\beta){\bf
{u}}_3 \ + \ \sin(\beta){\bf {u}}_4 $. The one parameter group generated by
this generator transforms the angular momentum as, (appendix B)
\begin{equation}
2J(\sigma) ~ = ~ 2J -\sin^2(\sigma)\left\{ 2J - \bar{{\bf {\omega}}}\bar{ {\bf u}} 
{\bf {L u \omega}} \right\} + \sin(2 \sigma) \left\{ \bar{ {\bf {\omega }}} {\bf { L u
\omega}} \right\}
\end{equation}

For the choice of {\bf u} made, $\bar{ {\bf u}} {\bf {L u}} \ = \ - {\bf L} $
and putting $J_u \ \equiv \ \frac{ \bar{ {\bf {\omega }}} {\bf { L u \omega}} }{2} $,

\begin{eqnarray}
J(\sigma) & ~ = ~ & J  \ \cos(2 \sigma) + \sin(2 \sigma)  \ J_u \nonumber \\
& ~ = ~ & \sqrt{J^2 + J_u^2}  \ \cos( 2 \sigma - \delta ) ~~, ~~~~~ \delta ~ 
\equiv ~ \tan^{-1}(J_u/J) ~  \in ~ (-\pi /2 , \pi /2).
\end{eqnarray}

This is valid for all choices of $\beta$. Observe that $J_u$ depends on
$\beta$. Furthermore it satisfies,
\begin{eqnarray}
\frac{\partial^2 J_u(\beta)}{\partial \beta^2} ~ & = & ~ - J_u(\beta) 
~~~~~~~~~~~~~~~~~~~~~~~~~~~~~~~~~ \Rightarrow \nonumber \\
J_u(\beta) ~ & = & ~ J_u(0) \cos(\beta) + \frac{\partial J_u}{\partial \beta}(0)
\sin(\beta) ~~~~~~~~ \Rightarrow \\
(J_u)_{max} ~ & = & ~ \ \left\{ \ (J_u(0))^2 \ + \ \right.
\left. ( \frac{\partial J_u}{\partial \beta}(0) )^2 \ \right\}^{1/2} 
~~~~~ = ~~ \sqrt{ \ E^2 \ - \ J^2 \ } \nonumber
\end{eqnarray}

The last equality follows by explicit evaluation. Choosing $\beta$ to
maximize $J_u$ then implies that the square root in the equation for 
$J(\sigma)$ is just the energy $E$. Since for any given energy, we must have 
$J^2 \le E^2$, it follows that $J(\sigma)$ covers the entire range of possible 
$J$ values. 

From this it follows that we can always choose a $\hat{\sigma}$ such 
that $J(\hat{\sigma}) \ = \ J^{\prime}$ without passing through zero angular
momentum. This proves the claim. Hence there are precisely {\it two} families 
of trajectories each spanning the whole orientation class. This gives the
classification of trajectories.

One may view the $S^3$ as compactification of $R^3$ with the south pole as the
origin of $R^3$ and the north pole as the point at infinity. The removal of
$\Delta$ now corresponds to removal of an infinite line, say, the z-axis of
the $R^3$. It is immediate that the fundamental group is the group of
integers, $Z$, and therefore the basic trajectories winding around the z-axis 
belong to two orientation classes. In this case the fundamental group directly 
shows how the basic trajectories are naturally classified into the ``orientation 
classes". 

Consider now the Hamiltonian vector fields generating the three one-parameter 
subgroups of $SU(2), {\bf u}_1$ generates rotations
about the z-axis. Its orbit through any generic point avoids the
z-axis and the vector field remains complete (its integral curves range over the
full real line). The other two subgroups on the other hand have orbits 
{\it {necessarily}} cutting through the z-axis. These vector fields therefore
are necessarily incomplete (for the anyon case). The transformations 
generated by these vector fields do {\it{not}} form a group and the dynamical 
symmetry group for anyons is reduced from $U(2)$ to $U(1) \times U(1)$.

Thus the removal of the set $\Delta$ has altered the fundamental group, has
classified trajectories into orientation classes and has also reduced the
dynamical symmetry group form $U(2)$ to $U(1) \times U(1)$ due to the
incompleteness of vector fields. 

Note that while the symmetry group is reduced from $U(2)$ to $U(1) \times
U(1)$ for anyons, the rank has remained the same and therefore the 
two-anyon system 
continues to be integrable. The integral curves being circles, we have
integrability via action-angle variables and we get the exact spectrum from 
the semiclassical approximation. In fact one can follows the EBK quantization
procedure to reproduce the exact spectrum. 

\subsubsection{The case of many anyons}

The $N > 2$ case differs from the $N = 2$ case in many ways. The potential
symmetry group, $OSp(4N, R)$, (see appendix B) generating families of trajectories 
is lot more complicated and so is the fundamental group for the phase space of 
anyons. However the analysis of the previous section already suggests a strategy 
for obtaining a classification of the families of trajectories. A generic 
trajectory may be viewed as a collection of $\frac{N ( N - 1 )}{2}$ ellipses 
traced by the $\vec{r}_{ij} , i < j $. To get a convenient handle on the 
disallowed trajectories, define
$J_{ij} \ \equiv \ ( \vec{r}_{ij} \times \dot{\vec{r}}_{ij} )_z $. Clearly a
trajectory will cut through the set $\Delta$ if and only if $J_{ij}$ vanishes
for at least one pair of indices. We will term such a trajectory as degenerate. 
For anyons only non-degenerate trajectories are allowed. Note that $J_{ij}$
are constants of motion just as $J_i$ and $E_i \equiv \frac{1}{2}( r_i^2 +
p_i^2)$ \cite{pseudo} are. Each of these elliptical trajectories will have 
a sense of traversal ( $ \epsilon_{ij} \ \equiv \ $ sign of $ J_{ij}$ ). Thus
a basic trajectory of $N$ particles has an associated set: \{$\epsilon_{ij}$\}.
Exactly as in the previous sub-section, the set of non-degenerate trajectories 
can be classified by these `orientations'. Provided we can exhibit trajectories 
which realize all possible choices of $\epsilon_{ij}$, we will have 
{\it at least} $2^{N ( N -1 )/2}$ families for anyons and precisely 
as many if every member of a class is connected to every other member of the 
same class by a symmetry transformation without crossing $\Delta$. It is easy
to show that every choice of $\epsilon_{ij}$ is realized by considering
concentric circular trajectories for each of the particles with various
possible signs for the angular momenta $J_i$ and various possible ordering of
the radii of the circular trajectories. For oscillator we will have precisely 
one family if $OSp(4N, R)$ acts transitively on the constant energy sphere, 
$S^{4N -1}$. 

The proof of transitivity in the case of oscillator is most directly obtained
by noting that $OSp(4N, R)$ is isomorphic to $U(2N, C)$ and it is well known
(and easy to see) that the coset space $U(2N,C)/U(2N -1, C)$ is diffeomorphic
to $S^{4N -1}$. This immediately shows that for the oscillator, there is
precisely {\it one} family of periodic trajectories.  The counting of 
families is more tricky in the case of the anyons. Let us see the symmetry 
reduction in a slightly different manner.

Let $\Delta^{\prime}$ denote the set of all degenerate trajectories i.e. at
least one of the $J_{ij}$'s is zero. Note that $\Delta$ and $\Delta^{\prime}$
are different, the former being a proper subset of the latter. Suppose we find
the subgroup of symmetries which leave $\Delta^{\prime}$ invariant, then the
same subgroup will also leave the set of all non-degenerate trajectories
invariant. Since we are looking for a group, we can consider the
infinitesimal action. Let $J \ \equiv \ \prod_{i < j} J_{ij}$. Clearly, $J =
0$ characterizes the set $\Delta^{\prime}$. If we have a degenerate
trajectory with two or more $J_{ij}$'s zero, then every infinitesimal action
will keep within the set of degenerate trajectories. If only one $J_{mn}$ is
zero then infinitesimal action will act invariantly if and only if $J_{mn}$
is invariant. Of course we could have degenerate trajectories with different
angular momenta being zero and hence for the invariance of $\Delta^{\prime}$ 
it is necessary that the generators of the subgroup must leave each of the 
$J_{ij}$ 's invariant {\it whenever $J_{ij}$ itself is zero}. This subgroup 
is identified in the appendix B. There are two forms of generators. Any
generator with all the block matrices being the same {\bf u} matrix with
${\bf u} \ \in \ OSp(4,R)$ will leave all the J's invariant. 
These transformations act on the center-of-mass variable alone and
constitute the expected $OSp(4,R)$ symmetry. In addition, generators with only
diagonal blocks being the same {\bf u} and all the off-diagonal blocks being
{\bf 0} also leave the J's invariant provided ${\bf {u \ = \ u}_1}$ or ${\bf
u}_2$ (or a linear combination thereof). These are just the total Hamiltonian 
and the total angular momentum ($\sum_i J_i$) which generate time evolutions 
and common rotations of all the positions and momenta. These are the only 
subgroups leaving the set of degenerate (and non-degenerate) trajectories 
invariant. The vector fields corresponding only to these will remain complete. 
Thus we see that the symmetry is reduced to $OSp(4,R) \times O(2) \times O(2)$. 
This is identical to the symmetry in the case of two anyons.

It is apparent from the above discussion that pseudo-integrability of anyons
really means that not all the Lie algebra level symmetries exponentiate to
Lie group symmetries. One should distinguish between ``infinitesimal
integrability" and integrability. A general discussion is given in the 
sub-section 5.3 . 

This purely classical analysis has already explained the qualitative result that
for anyons only the total energy and the total angular momentum are the
conserved quantities (good quantum numbers). Since we also have a
classification of periodic trajectories we will take the next step of
attempting semiclassical approximation to see what further information can be
obtained.

\subsection{Semiclassical Spectrum of anyons}

To develop a semiclassical approximation the central quantity of interest is 
the propagator defined as,
\begin{eqnarray}
G(E + i\epsilon) & \equiv & Tr( \frac{1}{E - H + i \epsilon} ) \nonumber \\
& = & \sum_{n} (E - E_n + i\epsilon)^{-1} \\
& = & \sum_{n}  ~ {\cal P} ( (E - E_n)^{-1} )  - i \pi \sum_{n} ~\delta (E - E_n). \nonumber 
\end{eqnarray}

In the last equality, ${\cal P}$ denotes the principle value while the second 
term is of course the density of states. Defining $\tilde{G} \equiv i G$ allows 
us to write the density of states, $g(E)$ as ,

\begin{eqnarray}
g(E) & = & \frac{1}{2\pi} ~ \{ \tilde{G}(E + i\epsilon) + \tilde{G}^{*}(E - i \epsilon)
~\} \nonumber \\
 & = & \frac{1}{\pi} ~ Re\{ \tilde{G}(E + i\epsilon) ~\}
\end{eqnarray}

The derivation of the semiclassical trace formula begins with the definition
of the propagator, $G(E)$, with trace operation expressed via a path integral
representation. The trace operation combined with the stationary phase
approximation (SPA) gives the propagator as a sum over periodic orbits (in the
phase space), of terms with an amplitude given by a suitable Van-Vleck 
determinant times a phase whose exponent is the value of the `action', 
$\int {\bf {p\cdot dq}}$ around the periodic orbit together with a 
contribution from the Maslov indices. This works very well for systems
having only isolated periodic orbits. For systems with symmetries, as for 
the oscillator or anyons, the periodic orbits come in continuous families 
and a modification is needed. 

In cases with symmetries, the sum over orbits gets replaced by a sum over 
families of orbits together with a measure factor. If the family arises due 
to a symmetry, as is usually the case, the amplitude and the phase is same
for all members of the family and as such can be computed from any one
member. This is discussed by Littlejohn et al \cite{littlejohn} in detail. 
The origin of families of periodic orbits in the case of oscillator is of course
the $OSp(4, R)$ dynamical symmetry and for anyons essentially the same 
transformations generate families of trajectories. This is already
discussed in the previous sub-section. 

A system could also admit the so called ``diffractive" orbits. Generically 
these are orbits which are closed in the configuration space but not in the 
phase space (or that the Hamiltonian flow is discontinuous)\cite{sieber}. 
Within the framework of so called ``uniformly continuous approximation" 
\cite{uniform}, these lead to further contributions to the semiclassical 
propagator. The anyon system presents us with both these features .

In the context of semiclassical approximation, one point needs to be noted.
Observe that in terms of the $\vec{\pi}_i , \vec{r}_i$ coordinates 
there is no explicit reference to $\alpha$ and $\Delta$ is also precisely the 
set on which one gets $\delta$-function singular Poisson brackets with $\alpha$ appearing as 
the coefficient. If one did inverse Legendre transformation to get a Lagrangian
one would not get the $\alpha$ dependent term, but of course the configuration
space will have the $\Delta$ removed. Using $\vec{p}_i, \vec{r}_i$ variables
and then doing inverse Legendre transformation will produce the Lagrangian
with the $\alpha$ dependent total derivative term. Now the removal of
$\Delta$ is made explicit by the $\alpha$ dependent term. It is this explicit
form which is convenient for computation of action for periodic orbits.

The action for the oscillator is equal to $(2\pi E)/\omega$ and the action for 
anyons is the same as that for the oscillator except for the 
contribution from the total derivative term in the Lagrangian. Its value is 
$\pm 2\pi \alpha$ with sign depending on the orientation class(es). We have
already seen that for $N = 2$ we have precisely $2$ orientation classes while
for general $N$ we have {\it{at least}} $2^{N(N-1)/2}$ classes. 

Are there ``diffractive contributions"? To appreciate this, let us consider 
the two anyon case in more detail. 
The propagator can be computed exactly from the known exact spectrum 
\cite{bhaduri}.  The exact spectrum is given by,
\begin{equation}
E_{n,j} ~ = ~ \hbar\omega ( 2n + | j - \alpha_{q} | + 1 ), ~~ n \ge 0,~~  j \in Z.
\end{equation}

From this the partition function is obtained as,

\begin{equation}
Z(\beta) ~ = ~ \frac{\cosh(~ \beta\hbar\omega~ (\alpha_{q} - 1)~ ) +
\cosh(~ \beta\hbar\omega\alpha_{q}~ )}{2 \sinh^2(\beta\hbar\omega)},
\end{equation}
and the exact density of states is obtained as:
\begin{eqnarray}
g(E) & \equiv & \frac{1}{2\pi i} ~ \int_{\epsilon - i \infty}^{\epsilon + i
\infty} e^{\beta E} Z(\beta)  \nonumber \\
& & \nonumber \\
& = & \frac{E}{(\hbar\omega)^2}\{ 1 + \sum_{k \ge 1}\{ 
\cos(\frac{2\pi k E}{\hbar\omega} + 2\pi k \alpha_q ) + 
\cos(\frac{2\pi k E }{\hbar\omega} - 2\pi k \alpha_q ) \} \nonumber \\
&  & - \frac{1}{(\hbar\omega)} \sum_{k \ge 1}\{ 
2\alpha_q ~  \sin(2\pi k\alpha_q) ~  \sin(\frac{2\pi k E }{\hbar\omega}) \} \\
&  & + \frac{1}{(\hbar\omega)} \sum_{k \ge 1}\{ 
(-1)^k  ~ \sin(\pi k\alpha_q)  ~ \sin(\frac{\pi k E }{\hbar\omega}) \}. \nonumber 
\end{eqnarray}

Rewriting the products of sines as differences of cosines, we get
\begin{eqnarray}
g(E) & = & \frac{E}{(\hbar\omega)^2}  \nonumber \\
&  & + \frac{1}{\hbar\omega} ~ (~ \frac{E}{\hbar\omega} + \alpha_q ~)
\sum_{k \ge 1} ~ \cos(\frac{2\pi k E}{\hbar\omega} + 2\pi k \alpha_q )  ~ 
\nonumber \\ 
&  & + \frac{1}{\hbar\omega} ~ (~ \frac{E}{\hbar\omega} - \alpha_q ~)
\sum_{k \ge 1} ~ cos(\frac{2\pi k E}{\hbar\omega} - 2\pi k \alpha_q )  ~ \\ 
&  & - \frac{1}{2\hbar\omega} ~ 
\sum_{k \ge 1} ~ (-1)^k ~ \cos(\frac{\pi k E}{\hbar\omega} + \pi k \alpha_q ) 
\nonumber \\ 
&  & + \frac{1}{2\hbar\omega} ~
\sum_{k \ge 1} ~ (-1)^k ~ \cos(\frac{\pi k E}{\hbar\omega} - \pi k \alpha_q ) 
\nonumber  
\label{gofe}
\end{eqnarray}

{\bf {Remarks:}}

1. The partition function is manifestly invariant under $\alpha_q \rightarrow
1 - \alpha_q$ . The convergence of integrals in computing the density of states for 
$\alpha_q \in (0, 1)$ requires $\frac{E}{\hbar \omega}$ to be greater than or 
equal to 1. 

2. The first term above is the usual Thomas-Fermi term \cite{bhaduri}.	This
will be suppressed in expressions below. 

3. In the limit $\hbar \rightarrow 0$ keeping $E, \alpha_q$ fixed, the exact 
propagator  effectively looses all dependence on $\alpha_q$. This is also the 
semiclassical trace formula for the oscillator showing that the semiclassical 
approximation is exact for the oscillator. All the other terms are sub-leading.

4. For a comparison with semiclassical approximation, a different limit
is implied. One should use $\alpha_c = \hbar \alpha_q$
and the SPA will be done keeping $\alpha_c$ fixed. Then only the last two
terms will be sub-leading and the propagator will continue to have $\alpha_c$
dependence. This is the limit considered in the following.%

5. In the last two sub-leading terms the argument of cosine is half of that in 
second and the third term. As is well known, the argument of 
cosines is the classical action, $\int \vec{p}\cdot d\vec{r}$, around a 
classical orbit including possible contributions from ``Maslov indices". The 
last two terms are therefore suggestive of contribution from ``half orbits". 
Indeed if one takes a Fourier transform of $g(E)$ \cite{bhaduri} , then one 
sees a peak at 1/2 period from the last two terms. {\it One may therefore 
suspect a contribution from classical trajectories with half the period.} Do
we have such classical trajectories?

Recall that anyons are fundamentally defined quantum mechanically and all the
energy eigenfunctions of the system vanish on the set $\Delta$ of coincident 
points. We modeled the classical system by removing $\Delta$ and explored the 
consequences in the previous sections. In particular we just {\it {omitted}}
trajectories that could cut through $\Delta$.

From a purely classical point of view, though, this is a little unsatisfactory. 
One could legitimately ask just what happens to trajectories (e.g. $J = 0$ for 
$N = 2$) that attempt reaching the disallowed region? To decide this one has 
to {\it{extend}} the classical modeling by supplementing the equations of 
motions with a specified ``boundary" condition. The choice is helped by 
noting that the curl of the vector potential is a $\delta$-function which 
may be ``regulated" to stipulate the ``boundary condition". This analysis is 
given in appendix A. 

The result is that {\it {all the elliptical trajectories are identical to 
those of the oscillator and only the degenerate trajectories get modified. 
A degenerate trajectory must reflect back from the coincident point.}} 
Clearly such a trajectory will have half the period of the generic 
trajectories and hence we do have such ``half trajectories". While this 
can been seen directly in the oscillator confinement from the form of the 
density of states, it may also be seen in the numerical computation of 
the Fourier transform of the density of states of a particle moving on a 
two-dimensional disc in the presence of the flux line\cite{bhaduri}.

While classically there {\it are} half-period orbits, the inference that
they are manifested in the two sub-leading terms of eq.(\ref{gofe}), is
ambiguous. To see this let us express the propagator in a sum-over-orbits
form. To do this rewrite the the last two terms to resemble the second and
the third terms. 

The last two terms also have a $(-1)^k$ in the summation over $k$ and this can
be handled in two ways. 

(A) Separate these sums into even and odd integer sums. All the sums over $k$ 
being geometric series can be done explicitly 
($i\epsilon$ needs to be added). Defining ${\cal{E}} = E/(\hbar\omega)$, we get,
\begin{eqnarray}
\hbar\omega g(E) & = & \cal E ~ + \nonumber \\
& & \frac{\frac{1}{2} ~ ({\cal E} + \alpha_q - \frac{1}{2} ~ )e^{2\pi i ( {\cal E}
+ \alpha_q )} + \frac{1}{4} e^{i\pi ({\cal E} + \alpha_q) } }{1 - e^{2\pi i \alpha_q
({\cal E} + \alpha_q )}} ~ + ~ \mbox{C. C.} \\
& + & \frac{\frac{1}{2} ~ ({\cal E} - \alpha_q + \frac{1}{2} ~ )e^{2\pi i ( {\cal E}
- \alpha_q )} - \frac{1}{4} e^{i\pi ({\cal E} - \alpha_q) } }{1 - e^{2\pi i \alpha_q
({\cal E} - \alpha_q )}} ~ + ~ \mbox{C. C.} \nonumber 
\end{eqnarray}

Here C. C. means complex conjugation.  Now we can ``read-off" $G(E + 
i\epsilon)$ and get (suppressing the Thomas-Fermi term),

\begin{eqnarray}
(-2\pi i)^{-1}  \hbar\omega G(E + i\epsilon) & = & \frac{1}{2} ~ 
\{ {\cal E} + \alpha_q - \frac{1}{2} ( 1 - 
e^{-i\pi ( {\cal E} + \alpha_q + i\epsilon ) } ) ~ \} ~~ \{
\frac{ e^{2\pi i ( {\cal E} + \alpha_q + i\epsilon)}}{ {1 - e^{2\pi i  
({\cal E} + \alpha_q + i\epsilon)}} } \}  + \nonumber \\
& & \frac{1}{2} ~ \{ {\cal E} - \alpha_q + \frac{1}{2} ( 1 - 
e^{-i\pi ( {\cal E} - \alpha_q + i\epsilon ) } ) ~ \} ~~ \{
\frac{ e^{2\pi i ( {\cal E} - \alpha_q + i\epsilon)}}{ {1 - e^{2\pi i 
({\cal E} - \alpha_q + i\epsilon)}} } \}.
\end{eqnarray}

Noting that the second group of braces is sum of a geometric series:
\begin{equation}
\frac{ e^{2\pi i ( {\cal E} \pm \alpha_q + i\epsilon)}}{ {1 - e^{2\pi i 
({\cal E} \pm \alpha_q + i\epsilon)}} }  =  
\sum_{k \ge 1} ~ e^{2\pi i k (\frac{E}{\hbar\omega} \pm \alpha_q + i\epsilon)},
\end{equation}
we see that the propagator {\it{is}} expressed in a form suggestive of a 
sum over periodic orbits. The first group within the braces being the 
``amplitude" while the second group being contributions from multiple 
traversals of a basic periodic orbit. The two terms can be seen to come 
from families of elliptical orbits going clockwise and anti-clockwise 
and that there are {\it {no}} term corresponding to degenerate ellipses or half 
orbits. Note that the terms in the ``amplitudes" other than ${\cal E} \pm
\alpha_q $ are sub-leading relative to ${\cal E} \pm \alpha_q $. This 
leading part
of the ``amplitudes" can also be seen to come from POT in presence of
continuous families of periodic orbits \cite{littlejohn}. The ``amplitudes" 
however are complex. For $\alpha_q = 0 $ there is cancellation of these 
sub-leading terms and one recovers the exact result for the oscillator. It 
appears that though the system is integrable, the trace formula does 
{\it{not}} give exact propagator. 

The poles in $G(E)$ come from the two terms and are given by:
\begin{eqnarray}
E_+(n_+) & = & \hbar \omega ( n_+ - \alpha_q ) ~~~~~~ n_+ ~ \ge 2 \\ 
E_-(n_-) & = & \hbar \omega ( n_- + \alpha_q ) ~~~~~~ n_- ~ \ge 1 .
\end{eqnarray}

The residues at these poles ( ${\cal E} \pm \alpha_q = n_{\pm}$ ) are $n_{\pm}$
if $n_{\pm}$ is {\it even} and $n_{\pm} - 1$ if $n_{\pm}$ is {\it odd}. These
residues of course give the degeneracy. {\it Note that the sub-leading terms are
important for this.} 

It is easy to see that the locations of all the poles can be re-expressed in 
the form given by the exact spectrum. The degeneracies at the poles also match 
exactly as expected. 

(B) We can also write $(-1)^k cos(k\theta) = cos(k(\theta \pm \pi))$.
Proceeding exactly as in the case (A), we can ``read-off" $G(E + i\epsilon)$
and get, 
\begin{eqnarray}
\left( \frac{-\hbar\omega}{2\pi i} \right) G(E + i\epsilon) & = & \frac{1}{2} 
\{ {\cal E} + \alpha_q \} \{
\frac{ e^{2\pi i ( {\cal E} + \alpha_q + i\epsilon)}}{ {1 - e^{2\pi i  
({\cal E} + \alpha_q + i\epsilon)}} } \} ~ + 
~ \frac{1}{2} \{ {\cal E} - \alpha_q \} \{
\frac{ e^{2\pi i ( {\cal E} - \alpha_q + i\epsilon)}}{ {1 - e^{2\pi i 
({\cal E} - \alpha_q + i\epsilon)}} } \} \nonumber \\
& & - \frac{1}{4} ~ \{
\frac{ e^{\pi i ( {\cal E} -1 + \alpha_q + i\epsilon)}}{ {1 - e^{\pi i 
({\cal E} -1 + \alpha_q + i\epsilon)}} } \} ~~~ + 
~~~ \frac{1}{4} ~ \{
\frac{ e^{\pi i ( {\cal E} + 1 - \alpha_q + i\epsilon)}}{ {1 - e^{\pi i 
({\cal E} + 1 - \alpha_q + i\epsilon)}} } \}. \\
& & \nonumber
\end{eqnarray}

These have poles which are subsumed by the poles from the first two terms. 
The overall pole structure and residues of course are exactly same as before as 
they should be. 

The exponents in the last two terms however do {\it not} look like the action
integral. A reflecting trajectory will not receive a contribution from the
$\alpha_q$ dependent total derivative term. The reflecting trajectories will
also form a single family of trajectories and thus should give only one term
and not two terms. The last two terms can be combined and the propagator may
be re-expressed as
\begin{eqnarray}
\left( \frac{-\hbar\omega}{2\pi i} \right) G(E + i\epsilon) & = & 
\sum_{k \ge 1} ~ \left[ ~ ( \frac{{\cal E} + \alpha_q }{2} ) ~ 
e^{2\pi i k ({\cal E} + \alpha_q + i\epsilon)}  ~ + ~ 
( \frac{{\cal E} - \alpha_q}{2} ) ~ e^{2\pi i k ( {\cal E} - \alpha_q + i\epsilon)}
 \right. \nonumber \\
& &  \left. ~ + ~ ( \frac{-i \ \sin(k \pi \alpha_q)}{2} ) ~ e^{i \pi k 
({\cal E} - 1)} ~ \right]. \\
& & \nonumber
\end{eqnarray}

The propagator now does have form indicating a contribution from the half
period orbits including a contribution from a Maslov index due to reflection.
However it has a complex amplitude with a dependence on $\alpha_c$ and the
multiple traversals index $k$. 

Thus the exact propagator can be expressed in two different forms mimicking
the trace formula with and without the half period trajectories. Although the
phases are as expected, the amplitudes are not. The ``amplitudes" are
complex in general and also have terms with different orders of $\hbar$. 

The different orders of $\hbar$ seen can be understood in the context of
trace formula as due to {\it families} of periodic trajectories and not due
to any higher order corrections to the SPA. When periodic trajectories come in
continuous families, the number of Gaussian integrations is reduced since the
integration over the family must be done exactly \cite{littlejohn} and this
increases the powers of $\hbar$ in the denominator. The elliptic trajectories
form a three parameter family while the half trajectories form a one
parameter family which explains the difference in the powers of $\hbar$. 

While the powers of $\hbar$ can be understood, the precise matching of the
amplitudes from the POT does not follow. One may conclude from this that
{\it {the exact propagator is not of the form implied by POT and that
additional contributions are necessary}}. The ``half period" peaks in the 
Fourier transform is a property of the exact propagator and need not
imply presence of ``half orbits" in a POT, a ``diffractive" contribution
may also generate such a peak. 

This analysis of the exact propagator shows that even though the system
is integrable, the propagator is not expressible as a POT sum, with or
without ``half orbits". In the light of the work on ``diffractive"
contributions cited above, the exact propagator seems to need such
contributions. However we have not done a precise demonstration that the
exact propagator {\it {can}} be obtained by inclusion of ``diffractive"
contributions. This is beyond the scope of present review. 

Needless to say, the structure of ``half trajectories" will be more
complicated for $N > 2$. One needs to develop/adapt the machinery of
diffractive contributions to this case to strengthen a semiclassical
demonstration of the non-linearly interpolating eigenvalues. 

Observe that {\it {even without}} the knowledge of precise number of
families, we have precisely $2^{N(N - 1)/2}$ sums in the sum over
periodic trajectories. This is because, the contribution of the action
integral depends only on the orientation class and not on further possible
subclasses of trajectories. The measure factors therefore add up within each
orientation class and this is sufficient to get the semiclassical
eigenvalues. 

The sum over families in the trace formula contains as many terms. To get the 
eigenvalues we need to look at only the action integral which differs from
that of  the 
oscillator only by the additional $\oint \frac{\alpha_c}{2} \ \sum_{i \ne j} 
\dot{\theta}_{ij}$ contribution. The oscillator's contribution to the action 
integral is of course just $\frac{2 \pi}{\omega} E$, where $E$ is the classical 
energy and $\omega$ is the oscillator frequency. 

The net $\alpha_c$ dependent contribution to the action integral is given by,
\begin{equation}
\frac{\alpha_c}{2} \oint \sum_{i \ne j} \dot{\theta}_{ij} ~ = ~ 2 \pi \alpha_c \sum_{i
< j} \epsilon_{ij} .
\end{equation}
Here $\epsilon_{ij}$ is the sign of traversal of $\vec{r}_{ij}$ and is $\pm
1$. The $\sum_{i < j}$ is trivial to evaluate. \\

Since the classical trajectories are identical to those of the oscillator, we
will have the same Maslov indices which incorporate the usual zero point
energy namely, $N(N-1)$ for $2N$ dimensional oscillator. Including this
contribution, the semiclassical energies are given by :

\begin{equation}
E_n \{\epsilon_i\} ~ = ~ \hbar\omega \left[n ~ + ~ 
\alpha_q \sum_{i < j } \epsilon_{ij} ~ + ~ N(N-1) ~ \right]~~~~~~~~~~ n ~ \ge 0.
\end{equation}

The coefficient of $\alpha_q$ takes all possible integer values from 
$ - N (N -1)/2 \ $ to $ \ N (N -1)/2 \ $ in steps of 2. The extreme
values correspond to all the particles traversing the same way and represent
the contribution of ``collective" motion shown in \cite{pseudo}. The
remaining values represent contributions of ``relative" motion. These are
new eigenvalues albeit at the leading $\hbar$ level. Numerical spectra
available for $N = 3, 4$ show eigenvalues matching with the above at the
$\alpha_q \ = \ 0$ and $\alpha_q \ = \ 1$ \cite{numerical}. These numerical 
eigenvalues however are non-linearly interpolating while our semiclassical 
eigenvalues are linearly interpolating. 

In reference \cite{numerical} also it was conjectured that there will be
linearly interpolating eigenvalues with various slopes based on a different
semiclassical argument. We have obtained these eigenvalues by direct
application of the periodic orbit theory and proved their existence for all
$N$. 

Observe that the $\alpha_q$ dependence in the eigenvalues ( locations of 
poles of $G(E)$ ) arises only from the explicit $\alpha_c \equiv \alpha_q 
\hbar$ in the action. This therefore must be at the most linear in $\alpha_q$.
Since $\alpha_q$ is fixed, the semiclassical limit of $\hbar \rightarrow 0 $ 
implies $\alpha_c \rightarrow 0$. Thus one may at the most see linear
$\alpha_c$ dependence in the leading approximation which could however be
indicative of exact non-linearly interpolating eigenvalues. 

Thus the leading level semiclassical propagator (with POT only) already 
indicates the presence of eigenvalues such that $E(\alpha_q = 1) - 
E(\alpha_q = 0)$ takes all possible integer values between $\pm N(N-1)/2~$ in 
steps of 2. 

As there does not appear to be any scope for going beyond the leading SPA 
(all intermediate exponents of phases are quadratic), it may be
conjectured that diffractive contributions are needed to improve these
semiclassical eigenvalues.

\subsection{Pseudo-integrability}

In the previous sections we encountered two important features, one
related to the fundamental group and one related to symmetry reduction due
to incomplete vector field. Both were caused by the same source, namely
removal of coincident points implying topologically non-trivial phase
space. Both are relevant for periodic orbit theory in a general way and a
few general remarks are in order. 

The stationary phase approximation to the propagator naturally leads to
periodic orbits in the phase space. These orbits may be isolated or come
in continuous families or both. The continuous families may be generated
by a full group of symmetries or by only a subset of symmetry
transformations. Since each periodic orbit is also a map of $S^1$ to the
phase space $\Gamma$, clearly every orbit must belong to one and only one
homotopy class of the fundamental group, $\pi_1(\Gamma)$, of the phase
space. If an orbit is a member of a continuous family then the entire
family must belong to a single homotopy class. Note that a given homotopy
class may contain no orbit (solution of equation of motion) or several
isolated orbits and/or several families of orbits. But a family can not
spill over two distinct homotopy classes. In sub-section 5.1 above, we saw 
precisely the splitting of a single basic family for oscillator into two basic
families for anyons because of the non-trivial fundamental group. 
Multiple traversals of course belong to different homotopy classes and are
explicitly summed over. Thus a non-trivial $\pi_1$ may (but not
necessarily) provide an obstruction to a symmetry. How exactly may such an
obstruction manifest itself? For this we have to consider vector fields
generating symmetries. 

Recall \cite{abraham} that every function on the phase space generates
{\it {infinitesimal}} symplectic diffeomorphisms (canonical
transformations) via its corresponding (globally) Hamiltonian vector
field. The Lie algebra of such  vector fields is isomorphic to the Poisson
Bracket (PB) algebra of functions on $\Gamma$.  However such infinitesimal
transformations exponentiate to give a one parameter group of
transformations {\it {only if}} the vector field is {\it { complete}} i.e.
the integral curves of the vector field can be extended so as to have
these as a map from the full $R$. Only complete vector fields---and this
is a global statement---give rise to groups of symmetries. (A
corresponding quantum mechanical statement for continuous symmetries is
contained in the Stone's theorem: every one parameter group of unitary
transformations is generated by a self-adjoint operator and conversely.)
Since we usually want classical symmetries to be reflected at the quantum
level with observable generators we have to have {\it {groups}} of
symmetries and hence complete vector fields. 

The criterion of integrability in terms of vanishing PB's guarantees
only the existence of {\it infinitesimal symmetries} which is of course a 
prerequisite. 
An ``infinitesimally integrable" system may thus be: (a) integrable via
action-angle coordinates if the vector fields are complete and the integral
curves are closed; (b) integrable, but not by action-angle variables if the
vector fields are complete but only a subset of these have closed integral
curves and (c) partially integrable if only a subset of vector fields are
complete.  The ``pseudo integrability" property of anyons pointed out in ref.
\cite{pseudo} falls in the category (c). As explained in the appendix B, apart 
from the Hamiltonian only the total angular momentum has a complete vector 
field and hence is the only quantum number that survives. 

A non-trivial fundamental group by itself however does not imply
possibility of incomplete vector fields. For, on a compact manifold all
non-singular vector fields are complete\cite{abraham} and it can of course 
have a non-trivial $\pi_1$. In many cases constant energy surfaces are compact
and one does not have to worry about incomplete vector fields. For the
anyons however removal of the set $\Delta$ makes the constant energy
surface non-compact which admits possibility of incomplete vector fields
and corresponding loss of symmetry. In the above, we have seen
manifestation of all these features. 


\section{Semi-classical analysis of the ground state}

The semiclassical analysis of the last section is a general one for all
the states of the N-anyon system. However, the ground state of this system
deserves special attention, being somewhat unusual. As pointed out by
Chitra and Sen \cite{tf}, due to the level-crossing between pairs of
energy levels, the ground state consists of many pieces at different
$\alpha$ values. This is because states of different $J$ have lowest
energy at different values of alpha. In this section we present briefly a
semiclassical analysis of the many-anyon ground state via another
approach, which is motivated by the above observation. 

We start from the classical solutions of lowest energy, which are the
stable configurations (``classical ground states'') of the system. These
states are determined by minimizing the total energy of the system. The
assumption is then made that in the semiclassical regime, the quantum
ground state energy is given by the classical energy to the zeroth
approximation with corrections to first order given by quadratic, quantum
fluctuations about this value. 

Now the quantum ground state doesn't interpolate smoothly between the
extreme values of the statistical parameter. This behaviour can never be
captured if we try to find the ground state by simply minimizing the
classical Hamiltonian. In fact, such a step gives a ``ground state'' which
is completely independent of the angular momentum, and therefore of
$\alpha$. 

The rotational symmetry of the Hamiltonian also means that the angular
momentum of the system is conserved. To find the minima of the energy, one
can proceed in two steps: first minimize the Hamiltonian for a fixed
angular momentum and then minimize the resulting energy with respect to the angular
momentum. In the second step, one could carry out the minimization over
{\it quantized} values of the angular momentum. This method, as we shall
shortly see, retains the dependence on angular momentum of the lowest
energy, even at the classical level and also has some features observed
in the Thomas-Fermi approach to the ground state \cite{tf}. 

\subsection{Classical ground states}

\begin{figure}[h]
\centerline{\epsfxsize 15cm
            \epsfbox{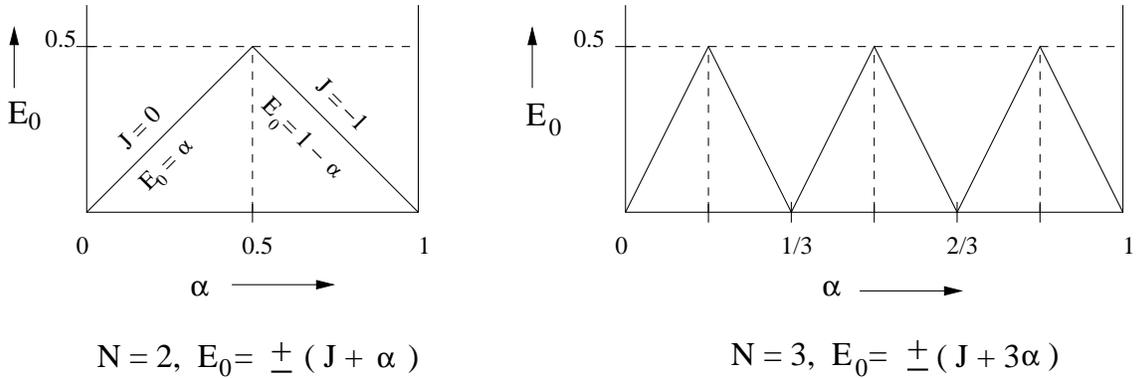} }
\caption{\sl Classical ground state energy of two and three
 anyons as a function of $\alpha$}
\label{fig:anyon}
\end{figure}

The energy of N free anyons in a harmonic confinement is represented by the
classical hamiltonian
\begin{eqnarray}
H = \frac{1}{2} \sum_{i}^N \left(({\bf p}_i - {\bf a}_i)^2 + r_{i}^{2}
  \right), \nonumber \label{hamanyon}\\
{\rm where~~~}
{\bf a}_i = \alpha \hat{\bf z} \times
\sum_{j \neq i} \frac{{\bf r}_{ij}}{r_{ij}^2}.
\end{eqnarray}
Constancy of angular momentum is applied as a constraint, 
\begin{eqnarray}
C \equiv \sum_i^N j_i - J ~~~~~
\mbox {where}~~~~
j_i =  x_i p_{y_i} - y_i p_{x_i}.
\end{eqnarray}
Introducing the Lagrange multiplier $\lambda$, the constrained
function
\begin{eqnarray}
F = H-\lambda C, \label{eqf}
\end{eqnarray}
must be minimised. That is, 
$\delta F=0$ gives the extrema which will be (local) minima if
$\delta^2 F > 0$.
The equations for equilibrium are thence derived to be
the following:
\begin{eqnarray}
p_{x_i} = (a_{x_i} - \lambda) y_i, \nonumber \\
p_{y_i} = (a_{y_i} + \lambda) x_i, \label{peq}
\end{eqnarray}
are conditions arising from the variation of $F$ with respect to the
momenta,
while variation with respect to the positions gives the equation
\begin{eqnarray}
(1 - \lambda^2){\bf r}_i = 0 .
\end{eqnarray}
Since this is true for all $i$, this fixes  the Lagrange
multiplier:  $\lambda = \pm 1$.
The constant angular momentum becomes 
\begin{eqnarray}
J = -\lambda \sum_{i}^N r_i^2 - \frac{\alpha}{2} N (N-1).
\end{eqnarray}
The equilibrium configurations are thus dynamic paths with non-zero
momentum and angular momentum, the positions being given by those
satisfying the above relation for a given angular momentum sector.
The energy in such a sector is given by
\begin{eqnarray}
E &=&  \sum_{i}^N r_i^2\nonumber \\
  &=& \pm ( J + \frac{\alpha}{2} N (N-1)).
\end{eqnarray}
To find the ground state energy $E_0$, one must now minimize this energy with
respect to $J$.
Since the main aim of the present exercise is to find a semiclassical
route to
anyon ground state energy, the above expression can be minimised with
respect to the {\em quantized}, integer values of $J$.
Now the term $\frac{\alpha}{2} N (N-1)$ can be expressed as a fraction
plus an integer: $\mu + n$ where $n \in \mathbb{Z}$. So the energy is
quantized
as $ E_0 = m + \mu$ where $m$ is an integer and $\mu$ is a fraction.
$E_0$ is evaluated and plotted as a function of $\alpha$
for different $N$ in figure \ref{fig:anyon}.

A remarkable feature of this picture is that the
the ground state for different values of $\alpha$ has a different $J$.
Note that the ground state energy for $N=2$ is exact, even 
quantum-mechanically.  For higher $N$, the ``sawtooth'' envelope of 
the minimum energy over the whole range of $\alpha$ reflects the 
behaviour observed in the earlier numerical studies on the
quantum ground state energy \cite{numeral,tf}. Level crossing effects are
seen at the classical level itself, as different $J$ values contribute
to the ground state at different $\alpha$.
Quantum corrections  could be expected to smooth out this graph.

\subsection{Towards Semiclassical corrections}

Quantum mechanically,
the ground state energy will be corrected by
the zero-point energy due to fluctuations about the classical minimum
energy configuration. Now we can make the assumption that the potential is
approximately harmonic near the minimum positions. The quantum
wave function is localized around these points. To calculate the
corrections to the energy, one can expand the potential
as a Taylor series about the minimum. The zeroth order term
is of course the classical ground state energy. The first order term
is zero and the second order term is quadratic in the fluctuations.
The matrix of second derivatives of the potential, or the
Hessian, can be diagonalised in the basis of the normal modes of the
system. Then the second order term for the expanded potential is just a
set of oscillators in the normal modes, whose squared frequencies are
the eigenvalues $\omega_i^2$ of the Hessian. These oscillators can
be quantized to give corrections
$\Delta E = 1/2 \sum_{i=1}^{N} \hbar \omega_i$, to
first order in $\hbar$. This of course is valid only if one can
neglect the higher order terms in the expansion of $V$ (see, for
example, Rajaraman\cite{raja}, chapter 5).
One hitch in this  program is the occurrence of {\em zero modes}:
 when any one of the $\omega_i$'s is
zero, then this approximation is no longer valid. This happens when
the configuration is one of neutral stability, i.e. there exist
directions of symmetry in the model. Perturbations along the symmetry
direction do not change the energy of the system. The quantum energy
eigenfunctions cannot be said to be localized near the minimum: they
will spread along the flat ``valley".
Zero modes can be dealt with in various ways. One of the most
straightforward methods is discussed in chapter 8 of
\cite{raja}.
The basic idea is to work in a system of coordinates where the
symmetry directions get separated when calculating the quantum
corrections.
The method we will adopt is to perform a transformation in the
coordinates such that the direction of symmetry as the collective
coordinate becomes apparent and can be isolated. The Hamiltonian is
manifestly independent of this coordinate and the corresponding
conjugate momentum is conserved. Hence the fluctuations along this
direction can be ignored and the Hessian, being only in terms of the
remaining coordinates, picks up the contribution from fluctuations
along directions that are not unchanged by the symmetry transformation.
This is equivalent to transforming the Hessian to a basis in which it
looks block-diagonal, such that one block contributes the zero
eigenvalues alone and the others do not. This block can be separated out.
One then diagonalises the remaining part of the matrix.

In the N-anyon case, we work with the Lagrangian 
\begin{eqnarray}
L = \frac{1}{2}\sum_{1}^{N}\Dot{\bf r}_i^2 + \sum_{1}^{N} \Dot{\bf
r}_i . {\bf a}_i - V(r_i),
\end{eqnarray}
where ${\bf a}_i$ is the usual statistical gauge potential and
$V(r_i)$ is some confining potential.

Overall rotations  is a symmetry and motion along the rotation coordinate can be
treated as a collective coordinate and separated out. This 
was already seen in section 4: we 
perform the time-dependent transformation
\begin{eqnarray}
{\bf r}_i = R(\theta) \bs{\eta}_i
\quad\text{with}\quad \eta_{y_N} = 0 \label{rot}, \\
\text{where~~~}
R(\theta) = \begin{pmatrix} \cos\theta(t)& -\sin\theta(t)\\
                            \sin\theta(t)& \cos\theta(t)
            \end{pmatrix}.
\end{eqnarray}

The Lagrangian then becomes
\begin{eqnarray}
L = \frac{1}{2}\sum_{1}^{N} \left( \Dot{\bs{\eta}_i}^2
    + 2 \Dot{\theta} \bs{\eta}_i \times \Dot{\bs{\eta}_i}
    + \Dot{\theta}^2 \eta_i^2 \right)
    + \Dot{\theta}\sum_{1}^{N} \bs{\eta}_i \times {\bf A}_i
    + \sum_{1}^{N}\Dot{\bs{\eta}_i}.{\bf A}_i - V(\eta_i),
\end{eqnarray}
The conjugate momenta are defined by the equations
\begin{align}
&P_{x_i} = \Dot{\eta}_{x_i} - \Dot{\theta}\eta_{y_i} + A_{x_i},\\
&P_{y_i} = \Dot{\eta}_{y_i} + \Dot{\theta}\eta_{x_i} + A_{y_i},\\
&P_{x_N} = \Dot{\eta}_{x_N} + A_{x_N},\\
&P_{\theta} = \Dot{\theta}\sum_{i=1}^{N} \bs{\eta}_i^2
          + \sum_1^{N-1}\bs{\eta}_i \times\Dot{\bs{\eta}}_i
          - \frac{\alpha}{2}N(N-1).
\end{align}
It is convenient to perform a canonical
transformation to new canonical momenta
\beqn
\bs{\pi}_i = {\bf P}_i - {\bf A}_i.
\eeqn
Taking $\eta_{x_N} \equiv X$, the Hamiltonian is then
\begin{eqnarray}
H = \frac{1}{2}\sum_{1}^{N} {\pi}_i^2
    +\frac{1}{2X^2}\left( P_\theta + \frac{\alpha}{2}N(N-1)
 -\sum_1^{N-1}\bs{\eta}_i \times \bs{\pi}_i \right)^2 + V(\eta).
\end{eqnarray}
and constrained minimization gives
\begin{equation}
\begin{split}
&\pi_{x_i} = -\lambda\eta_{y_i}, \\
&\pi_{y_i} = \;\lambda\eta_{x_i}, \\
&\pi_{X} = 0,\\
&\bs{\nabla}_i V =  \lambda^2 \bs{\eta}_i.
\end{split}
\end{equation}
For harmonic confinement, $V = \frac{1}{2}\sum \eta_i^2$ and the Lagrange
multiplier gets fixed to $\pm1$, the choice depending on which
minimizes the energy. The equilibrium conditions can then be written
down as
\begin{equation}\begin{aligned}
\bs{\pi}_i =& \lambda \hat{\bf k} \times  \bs{\eta}_i, \\
J =& \lambda  \sum_i^N \eta_i^2.
\end{aligned}
\end{equation}
This means that the classical extrema lie on a $(2N-1)$-dimensional
hyper-sphere in coordinate space  whose radius is the square root
of the magnitude of the
 total angular momentum.
The energy is
\begin{equation}
\begin{split}
E_0(J) =& \sum_1^N \eta_i^2\\
       =& \lvert J \rvert\\
       =& \lvert P_\theta + \frac{\alpha}{2}N(N-1) \rvert
\end{split}
\end{equation}
which is the same as obtained before.
The advantage is that in these coordinates the Hessian will not
contain
the zero mode corresponding to rotations. The Hessian matrix, evaluated
at the equilibrium positions, in terms
of the coordinates $\eta_{x_i}, \eta_{y_i}, i = 1 \ldots N-1$, momenta
$\pi_{x_i}, \pi_{y_i}, i = 1 \ldots N-1$ and finally the coordinate
$X$ and momentum $\pi_X$ of the $N$th particle, looks like
\begin{equation}
M = \begin{pmatrix}
   \Delta + \zeta \zeta^T & 2\zeta & 0\\
   2 \zeta^T & 4 & 0 \\
  0 & 0 & 1
  \end{pmatrix}, \label{Hess}
\end{equation}
where
\begin{align*}
\zeta^T & = \frac{1}{X}\begin{pmatrix}
        \eta_{x_i} & \eta_{y_i}&  -\lambda\eta_{y_i}&  \lambda
\eta_{x_i}
        \end{pmatrix}, \\
\Delta & = \begin{pmatrix}
         \mathbb{I}\rvert_{2N-2} & \Omega\rvert_{2N-2}\\
         - \Omega\rvert_{2N-2} & \mathbb{I}\rvert_{2N-2}
         \end{pmatrix},\\
\Omega & = \begin{pmatrix} 0 & \mathbb{I}\rvert_{N-1}\\             -
\mathbb{I}\rvert_{N-1} & 0
         \end{pmatrix}, \text{~~the symplectic form in $2N-2$
dimensions}.
\end{align*}
 
This matrix must now be diagonalised to get a quadratic form in the
normal modes.
In this case, since the ground states are  not
static this matrix is in terms of all the
phase-space variables and not merely the coordinates.
The diagonalization must therefore preserve the symplectic structure of
the phase space. The details of this symplectic diagonalization
are given in appendix C.

We find that the matrix has one eigenvalue equal to $1$,
$2N-2$ eigenvalues equal to $2$, one eigenvalue equal to 
 $ 2\left( 1 +\frac{|J|}{X^2} \right)$
and finally, extra $2N-2$ {\it zero} modes.
The eigenvectors corresponding to these extra zero modes are tangential to
the equilibrium phase space coordinates. Displacement along these
directions preserves the $2N-1$ dimensional sphere of equilibrium, so
they correspond to displacements along this sphere. These zero modes
cannot however, be removed 
using similar techniques as used above. The main reason for this is
that they correspond to infinitesimal displacements along the $2N-1$
dimensional sphere mentioned above, which is however punctured at the
positions of coincidence of particles. The displacements cannot be
made large, and therefore these zero modes do not actually represent a 
global symmetry direction of the system.

This path to finding the exact ground states semiclassically seems to
face difficulties here, and the origin seems to be precisely the
analytic intractability of the non-trivial energy levels that will
descend through level crossings. 
                                
\section{Summary and Discussion}

We have presented in this review several issues related to the classical
and quantum mechanics of anyons.  Since several books and reviews dealing
with various aspects of anyon physics already exist, we have focussed on
aspects that have received little or no attention until now.  With this in
view, we have specifically focussed on the integrability properties of the
anyon system and semiclassical analysis of the spectrum. We have therefore
discussed two properties of the many anyon system: (1) The partial
separability of collective and internal degrees of freedom and (2) its
identification as a pseudo-integrable system. We have shown that the first
property explains the existence of the exactly known eigenvalues which is
some what uncommon for a generic many body system.  {\it This also shows
that the exactly known spectrum incorporates only a some what trivial
aspect of anyon dynamics}. The nontrivial aspects although partially
uncovered by numerical results are still remain elusive. 

The second property enables one to understand qualitatively the origin of
regular and irregular features in the spectrum of many anyons. We have
clarified and sharpened the meaning of pseudo-integrability.  The system
is {\it locally} identical to the isotropic oscillator in two dimensions
but not at the {\it global} topological level. It is the non-trivial
global topology that reduces the symmetry group from that of the
oscillator. 

We also considered application of the periodic orbit theory
to this locally trivial (integrable) but globally non-trivial system.  As
a by-product, we saw how the exact spectrum for two anyons is reproduced 
by semiclassical methods. We saw that while the dynamical symmetry is
reduced from $SU(2)$ to $U(1)$, the rank remained the same and hence
integrability property is preserved.

For $N \ge 3$, we reproduced the previously known exact eigenvalues
\cite{pseudo}. In addition, we obtained further {\it {new}} linearly   
interpolating eigenvalues. In the language of reference \cite{pseudo},
these correspond to the signature of `relative' dynamics. In this case the
symmetry reduction was drastic, from $OSp(4N, R)$ to $OSp(4,R) \times
O(2,R) \times O(2,R)$. The rank was reduced from $2N$ to $6$, destroying
the integrability property.

We have also discussed in detail the issue of ``half" trajectories and
pointed out their ambiguous role. Taking the view-point that ``half"
trajectories be excluded, and noting that there does not seem to be any
scope for computing higher order corrections to the semiclassical spectrum
with a possible non-linear dependence on $\alpha_q$, it seems that the
non-linearly interpolating eigenvalues are genuine quantum consequences
beyond what semiclassical analysis could give. Semi-classical analysis is
nevertheless sufficient to indicate the presence of these eigenvalues. 
Thus, the impact of pseudo-integrability of the anyon system on the classical
periodic orbit theory is quite non-trivial.


\begin{flushleft}
Acknowledgments:
\end{flushleft}

Several people have contributed to our understanding of the anyon physics. 
Specifically we acknowledge our debt to Radha Balakrishnan, G. Baskaran,
Rahul Basu, R.K. Bhaduri, M. Brack, Sudhir Jain, A. Khare, M. Krishna, J. Law,
R. Ramaswamy, Diptiman Sen and R. Simon. Thanks are due to Diptiman Sen
for useful comments on the manuscript.


\newpage
\section*{Appendix A : Regularised classical dynamics and reflecting orbits } 
\addcontentsline{toc}{section}{Appendix A : Regularised classical dynamics} 

Consider without loss of generality the case of two anyon in the relative 
coordinates. Generic orbits in the configure space, $R^2 - \{\vec{0}\}$,
are of course an ellipse and only a degenerate elliptical orbit attempts to pass 
through the origin. Noting that this system can be thought of as a charged 
particle in presence of a singular (``statistical") magnetic field 
at the origin, one may regularise the magnetic field or the flux to study the 
orbits and obtain the limiting behaviour to deduce ``boundary condition" at the
origin. 

To do this we imagine the relative Hamiltonian to be that of a particle in a
magnetic field along the z axis. The field is of course to be effectively
confined to a small disc around the origin. Now observe that for an axially
symmetric magnetic field along the z-axis, $\vec{B(r,\theta)} = B(r) \hat{k} $ 
where $r, \theta$ are the usual spherical coordinates in two dimensions,  
we may write the vector potential as,
\begin{equation}
\vec{A} = A_r \hat{r} + A_{\theta} \hat{\theta} 
\end{equation}
 
This implies,
\begin{equation}
B(r) = \partial_r A_{\theta} + \frac{A_{\theta}}{r} -\frac{1}{r} \partial_{\theta} A_r ,
\end{equation}
 
In a symmetric gauge, the vector potential is independent of the polar angle
and by choosing it to be divergence free one can set the radial component to
zero. 

The flux $\Phi(R)$, through a disc of radius $R$ is given by,
\begin{equation}
\Phi(R) \ = \ 2 \pi \int_0^R dr r B(r) ~~ = R \oint A_{\theta} d\theta \ = \ 
2 \pi R A_{\theta}(R).
\end{equation}

This implies,
\begin{eqnarray}
A_{\theta}(r) = \frac{\Phi(r)}{2 \pi r} = 
\frac{1}{r} \int_{0}^{r} dr^{\prime} r^{\prime} B(r^{\prime}) 
\end{eqnarray}

The choice, $\Phi(r) \ = \ 2 \pi \alpha ~ \forall \ r > 0$, gives the vector
potential used in the quantum mechanical calculation. It also implies the $r
B(r) \ = \ \alpha \delta(r)$ and this of course is the singular nature of the
magnetic field. Notice that smearing the $\delta(r)$ will not make the
magnetic field non-singular at the origin because of the explicit $1/r$.

For a regulated system one wants the magnetic field to be non-singular every
where and effectively confined to a disk of radius $\epsilon$ around the
origin. This requires the vector potential $A_{\theta}$ also to be non-singular and
therefore vanishing at the origin. If $A_{\theta}(r) \ \rightarrow \ c
r^{\beta}$ as $r$ approaches zero then, $B(r) \ \rightarrow \ c(\beta + 1)
r^{\beta -1}$. For a finite, nonzero $B(0)$ we must have $\beta \ = \ 1$ and
therefore $\Phi(r) \ \rightarrow \ 2 \pi c r^2$ near the origin. For $r \ge
\epsilon$ we still retain the flux to be $2 \pi \alpha$. Continuity at $r =
\epsilon$ then gives $c = \alpha / \epsilon^2$. Notice that this limiting
behaviour is fixed by the demand of non-singularity of the fields and as such
must be reflected in any explicit choice for the magnetic field. 

Thus our regulation involves choosing,
\begin{eqnarray}
\Phi(r) & = & 2 \pi \alpha ~~~~~~~~ \forall ~~ r \ge \epsilon \nonumber \\
& \rightarrow &  2 \pi \alpha \frac{r^2}{\epsilon^2} ~~~~ \mbox{as} ~~ r \ 
\rightarrow 0 
\end{eqnarray}

One could choose a uniform nonzero magnetic field inside the disk as an
explicit choice but it will not be necessary. Since we are interested in the
limiting behaviour of trajectories as $\epsilon$ is taken to zero, the
limiting behaviour of the flux is all that we need. 

Consider now the orbit equation for $\epsilon$ nonzero. Integrating
the equations of motion once using the two constants, energy $E$ and angular
momentum $\ell$, the orbit equations in $r, \theta$ coordinates become :
\begin{eqnarray} 
\dot{r} ~~~~~ & = &~~~~~\pm \sqrt{2E - r^2 {\dot{\theta}}^2 - r^2 }   \nonumber \\
r^2 \dot{\theta} ~~~~~ & = & ~~~~~ \ell - r A_{\theta}(r) \\
& & ~~~~~ \nonumber \\
& = & ~~~~~
\left\{ 
\mbox{$ 
\begin{array}{lcl} 
\ell - \alpha \frac{r^2}{\epsilon^2} & ~~~~~ & r ~ < ~ \epsilon \\ 
\ell - \alpha & ~~~~~ & r ~ \ge ~ \epsilon 
\end{array}
$} \right. \nonumber 
\end{eqnarray} 

There are three types of orbits possible: those which are fully inside the
disc, those which are fully outside the disc and those which go both inside
and outside the disc. In the limit of $\epsilon$ going to zero, the first type
of orbits are clearly irrelevant. It is easy to see that for the second type
of orbits one must have $\ell \neq \alpha$. These are insensitive to the flux 
in the limit and are thus identical to the orbits of the oscillator.

For the last type, an interior turning point is possible
only for $0 \le \ell \le \alpha$. The condition that such an orbit must also
have an exterior turning point limits $\ell$ to $\alpha$. We are interested in 
computing the change in the angular coordinate from the entry into the disk till 
exit from it. Explicit computation shows that the change in the angular coordinate 
goes to zero as $\epsilon$ goes to zero. Thus such an orbit {\it reflects} at 
the origin. Note that these are precisely the radial orbits.  

To summarize, a regulated classical modeling for anyons is generically
stipulated by giving the behaviour of the flux near the origin. It amounts to
cutting out a disk of radius $\epsilon$ and filling it up with non-singular
fields. The classical non-radial orbits then are exactly same as those of the 
oscillator except for the replacement $l \rightarrow l -\alpha$. The radial 
orbits reflect at the the origin and are termed {\em half orbits} since 
their period is half of that for the other orbits. 

\newpage
\section* {Appendix B : OSp(4N, R) dynamical symmetry and classification of trajectories} 
\addcontentsline{toc}{section}{Appendix B : OSp(4N, R) dynamical symmetry} 

In this appendix we collect together some of the well known facts about
the dynamical symmetry of the $n$ dimensional isotropic oscillator. 

The isotropic oscillator in $n$ dimensions has $OSp(2n, R)$ as the group
of dynamical symmetries. This is a group of $2n\times 2n$ order
matrices, ${\bf g}$, which are both orthogonal and symplectic. Denoting by
$\bar{{\bf g}}$ the transpose of ${\bf g}$, we have the defining equations: 
\begin{equation}
\begin{array}{llclcl}
& \bar{{\bf g}} ~  {\bf g} & ~  = ~ & {\bf I}_{2n} & : & (\mbox{orthogonal}) \nonumber \\
& \bar{{\bf g}}  ~ {\bf {\Omega}}  ~ {\bf g} & ~ = ~ & {\bf {\Omega}} & : &  (\mbox{symplectic}) \nonumber \\
\mbox{and with} & {\bf g} & ~ \approx ~ & {\bf I}_{2n} + \epsilon {\bf T} & & \mbox{generators $T$ satisfy} \\
& \bar{{\bf T}} & ~ = ~ & - {\bf T} & & \nonumber \\
& \bar{{\bf T}} ~ {\bf {\Omega}} & ~ = ~ & - {\bf {\Omega}} ~ {\bf T} & & \nonumber 
\end{array}
\end{equation}

Here ${\bf I}_{2n}$ is the identity matrix of order $2n$ while ${\bf
{\Omega}}$ is a suitable matrix defining the symplectic condition.  This will
be chosen below. 

The symplectic condition ensures that we have a (linear) canonical
transformation while orthogonality ensures that the Hamiltonian, 
 $ \sum_i (p_i^2 + q_i^2)/2$ , is invariant. It is easy to see that dimension 
of this group is $n^2$ while its rank is $n$. In fact one can show that the 
group $OSp(2n,R)$ is 
isomorphic to the group $U(n,C)$. Making an explicit choice of {\bf
{$\Omega$}} in a block form, one can obtain the block form for {\bf g} using
the symplectic condition. The real block matrices can be combined into 
complex block matrices. The orthogonality condition in terms of real matrices
then translates into the unitarity condition for the complex matrices
thereby proving the group isomorphism. This isomorphism immediately implies
that the ortho-symplectic group acts transitively on the constant energy sphere. 
This is used in the section 5.1 .

We are interested in finding the action of this group on the phase space via 
canonical transformations (symplectic diffeomorphisms). For notational 
convenience let us group the standard canonical variables together and denote 
them by $\omega_{\mu} ~, \mu = 1,2, ... , 2n,$ the first $n$ being
coordinates and the last $n$ being the momenta. The dual of the symplectic
form has components, $\Omega^{\mu \nu}$ given by the $\mu \nu$ th element of
the matrix ${\bf {\Omega}}$. With this notation, the PB of functions on the
phase space and the infinitesimal canonical transformations are given by,
\begin{eqnarray}
\{ F(\omega) ~,~ G(\omega) \} & ~ = ~ & \Omega^{\mu \nu} ~ \partial_{\mu} F
~ \partial_{\nu} G \nonumber \\
\delta_{\epsilon}\omega^{\mu} & ~ = ~ & \epsilon ~ \Omega^{\mu \nu} ~
\partial_{\nu}F(\omega)
\end{eqnarray}

Functions purely quadratic in ${\mathbb {\omega}}$ generate linear canonical
transformations and are also closed under the PB's and those which leave the
Hamiltonian invariant give the symmetry transformations.  Explicitly,
%
\begin{equation}
\begin{array}{lclclcl}
F & ~ \equiv ~ & \frac{1}{2} {\mathbb {\bar{{\omega}}}} ~ {\bf A} ~ {\mathbf {\omega}} 
& ~ , ~ & 
G & ~ \equiv ~ & \frac{1}{2} \bar{{\bf {\omega}}} ~ {\bf B} ~ {\mathbb {
{\omega}} }
~~~~~ \Rightarrow ~ \nonumber \\
\{ F , G \} & ~ = ~ & \frac{1}{2} \bar{{\bf {\omega}}} ~ 
( {\bf {A \Omega B - B \Omega A}} ) ~ {\bf {\omega}} & & & & \\
\end{array}
\end{equation}

The Hamiltonian corresponds to $ {\bf A} = {\bf I}_{2n} $. The Poisson
bracket of any $G$ with $H$ vanishes provided the matrix {\bf B} commutes
with ${\bf {\Omega}}$. Thus generic matrices defining quadratic functions are
real, symmetric matrices commuting with ${\bf {\Omega}}$. The matrices ${\bf
{\Omega B}}$ are then antisymmetric and provide an isomorphism of quadratic
functions to the generators ${\bf T}$ of the group $OSp(2n, R)$. 

Action of the one parameter group generated by a function $G$ is found from
the integral curves of the corresponding Hamiltonian vector field. These
curves are defined by the matrix equations, ($ {\bf T} \equiv {\bf {\Omega
B}} $)
%
\begin{eqnarray}
\frac{d {\mathbf {\omega }}(\sigma) }{d \sigma} ~ = ~ {\bf {\Omega B \omega}} \\
{\bf {\omega}} (\sigma)  ~ = ~ ( e^{\sigma {\bf {\Omega B}}} ) {\bf {\omega}}( 0
)
\end{eqnarray}

It is convenient to choose a particular grouping of the phase space
coordinates and corresponding choice of ${\bf {\Omega}}$.  Firstly let us
put $n = 2N$, relevant for the present context, so that the phase space is 
$4N$ dimensional. Arrange the coordinates and momenta as $x_1, y_1, p_{1x}, 
p_{1y} , .... , x_N, y_N, p_{Nx}, p_{Ny}$ and denote by $\omega _{i}$ the 
coordinates and momenta of the $i^{th}$ particle. The index $i$ now runs 
from 1, ..., N and each $\omega_i$ is a $4 \times 1$ matrix. Correspondingly 
we choose ${\bf {\Omega}}$ as a block diagonal matrix with N blocks and each 
block being $4 \times 4$ matrix ${\bf {\Lambda}}$. We choose,
\begin{equation}
{\bf {\Lambda}}  ~ = ~ \left( \begin{array}{cc}
{\bf 0}_2 & {\bf I}_2 \\ 
{\bf -I}_2 & {\bf 0}_2 
\end{array} \right)  
\end{equation}

One can choose a basis for $OSp(4N, R)$ as follows. Let us denote the
generators as ${\bf T}_i$, $i = 1,2, ... ,N$ and ${\bf T}_{ij}$ with $i < j$.
Each of these are expressed in the block form. The ${\bf T}_i$ are block
diagonal with a nonzero $4 \times 4$ matrix ${\bf u}_i$ as the $i^{th}$
block element. The ${\bf T}_{ij}$ have a matrix ${\bf v}_{ij}$ at the $i^{th}
$ row and $j^{th}$ column ($(ij)^{th}$ block) and ${\bf {-{\bar{v}}}}_{ij}$ 
at the $(ji)^{th}$ block.  In equations, 
\begin{eqnarray}
( {\bf T}_i )_{mn} ~ = ~ {\bf u}_i ~ \delta_{im} ~ \delta_{mn} \nonumber \\
( {\bf T}_{ij} )_{mn} ~ = ~ {\bf v}_{ij} ~ \delta_{im} \delta_{jn} -
~ {\bf {\bar{v}}}_{ij} ~ \delta_{in} \delta_{jm}
\end{eqnarray}

With these definitions it is easy to translate the conditions on the
generators in terms of the $4 \time 4$ matrices {\bf {u , v}} as:
\begin{equation}
{\bf {\bar{u}}}_{i} ~ = ~ - {\bf u}_{i} ~~~~~~~~~~ 
{\bf u}_{ij}{\bf {\Lambda}} ~ = ~ {\bf {\Lambda}}{\bf u}_{ij} ~~ ; ~~~~~~~~~~ 
{\bf v}_{ij}{\bf {\Lambda}} ~ = ~ {\bf {\Lambda}}{\bf v}_{ij} 
\end{equation}

Thus the {\bf u}'s generate $OSp(4, R)$ while the {\bf v}'s are required to
commute with ${\bf {\Lambda}}$. The number of independent {\bf u}'s is 4
while number of independent {\bf v}'s is 8 of which $4$ are `diagonal' and 
$4$ are `off-diagonal'. The dimension of $OSp(4N, R)$  is thus 
$~4N + 8N(N-1)/2~=~4N^2$. These independent matrices can be explicitly 
chosen in terms of $2 \times 2$ Pauli matrices and the identity matrix as:
\begin{eqnarray}
{\bf u}_{(1,2,3,4)} & ~ \sim ~ &  
\left( \begin{array}{cc}
i{\bf {\sigma}}_2 & {\bf 0}_2 \\ 
{\bf 0}_2 & i{\bf {\sigma}}_2 
\end{array} \right) ~,~ 
\left( \begin{array}{cc}
{\bf 0}_2 & {\bf I}_2 \\ 
-{\bf I}_2 & {\bf 0}_2 
\end{array} \right) ~,~ 
\left( \begin{array}{cc}
{\bf 0}_2 & {\bf {\sigma}}_1 \\ 
-{\bf {\sigma}}_1 & {\bf 0}_2 
\end{array} \right) ~,~ 
\left( \begin{array}{cc}
{\bf 0}_2 & {\bf {\sigma}}_3 \\ 
-{\bf {\sigma}}_3 & {\bf 0}_2 
\end{array} \right) ~;~  \nonumber \\
{\bf v}_{(1,2,3,4)} & ~ \sim ~ &  
\left( \begin{array}{cc}
{\bf {I}}_2 & {\bf 0}_2 \\ 
{\bf 0}_2 & {\bf {I}}_2 
\end{array} \right) ~,~ 
\left( \begin{array}{cc}
{\bf {\sigma}}_1 & {\bf 0}_2 \\ 
{\bf 0}_2 & {\bf {\sigma}}_1 
\end{array} \right) ~,~ 
\left( \begin{array}{cc}
i{\bf {\sigma}}_2 & {\bf {0}}_2 \\ 
{\bf {0}}_2 & i{\bf {\sigma}}_2 
\end{array} \right) ~,~ 
\left( \begin{array}{cc}
{\bf {\sigma}}_3 & {\bf {0}}_2 \\ 
{\bf {0}}_2 & {\bf {\sigma}}_3 
\end{array} \right) ~,~  \\
{\bf v}_{(5,6,7,8)} & ~ \sim ~ &  
\left( \begin{array}{cc}
{\bf {0}}_2 & {\bf I}_2 \\ 
-{\bf I}_2 & {\bf {0}}_2 
\end{array} \right) ~,~ 
\left( \begin{array}{cc}
{\bf {0}}_2 & {\bf {\sigma}}_1 \\ 
-{\bf {\sigma}}_1 & {\bf {0}}_2 
\end{array} \right) ~,~ 
\left( \begin{array}{cc}
{\bf {0}}_2 & i{\bf {\sigma}}_2 \\ 
-i{\bf {\sigma}}_2 & {\bf {0}}_2 
\end{array} \right) ~,~ 
\left( \begin{array}{cc}
{\bf {0}}_2 & {\bf {\sigma}}_3 \\ 
-{\bf {\sigma}}_3 & {\bf {0}}_2 
\end{array} \right) ~,~  \nonumber 
\end{eqnarray}

We are interested in one parameter groups generated by some element {\bf T}
of the Lie algebra. If {\bf T} satisfies: ${\bf T}^2 = -{\bf P}$ with {\bf P}
satisfying ${\bf P}^2 = {\bf P} \ , \ [ {\bf T} , {\bf P} ] = 0$, then it follows
that, 
\begin{equation}
e^{\sigma {\bf T}} \ = \ {\bf I}_{4N} - {\bf P } + {\bf P} \left[ cos(\sigma) + sin(\sigma)
{\bf T} \right] {\bf P}
\end{equation}

For the basis generators ${\bf T}_i$ we have {\bf P} to be block diagonal
with ${\bf I}_4$ as the $i^{th}$ block while for ${\bf T}_{ij}$ we have {\bf
P} to be the block diagonal matrix with ${\bf I}_4$ as the $i^{th}$ and the
$j^{th}$ blocks. Using these, the exponentials of basis generators can be 
evaluated to get the integral curves as, (with obvious notation)
\begin{eqnarray}
{\mathbf {\omega}}(\sigma) & ~ = ~ & 
\left[ ~ {\bf I}^{\prime}_i \ + \ {\bf I}_i \ \left\{ \ cos(\sigma) ~ + ~ sin(\sigma) 
\ {\bf T}_i \ \right\} \ {\bf I}_i ~ \right] {\mathbf {\omega}} (0) ~~~~ \mbox{and,} \nonumber \\
{\mathbf {\omega}}(\sigma) & ~ = ~ & 
\left[ ~ {\bf I}^{\prime}_{ij} \ + \ {\bf I}_{ij} \ \left\{ \ cos(\sigma) ~ + ~  sin(\sigma) 
\ {\bf T}_{ij} \ \right\} \ {\bf I}_{ij} ~ \right] {\mathbf {\omega}} (0)  
\end{eqnarray}

Therefore integral curves of the basis generators are periodic curves. Further,
the ${\bf T}_i$'s affect only the $i^{th}$ particle position and momenta while
the ${\bf T}_{ij}$ mix the $i^{th}$ and $j^{th}$ particles only. 

We need to study how the angular momenta $J_i$ and $J_{ij}$ vary under the
action of one parameter subgroups. Define the (block) matrices,
\begin{eqnarray}
({\bf L}_i)_{mn} \ & \equiv & \ {\bf L} \delta_{im} \delta_{mn} ~~~~~~~~
{\bf L} \ \equiv \ {\bf v}_7 \nonumber \\
({\bf L}_{ij})_{mn} \ & \equiv & \ {\bf L} \left\{ \ \delta_{mn} ( \delta_{im}
+ \delta_{jm} ) - \delta_{im} \delta_{jn} - \delta_{in} \delta_{jm} \
\right\} 
\end{eqnarray}

In terms of these matrices the angular momenta are given by:
\begin{eqnarray}
J_i \ & = &  \ \frac{1}{2} \ {\bf \bar{\omega} \ L}_i \ {\bf {\omega}} ~~~~~~~
= \ \frac{1}{2} \ {\bf \bar{\omega}}_i \ {\bf L} \ {\bf {\omega}}_i \nonumber
\\
J_{ij} \ & = & \ \frac{1}{2} \ {\bf \bar{\omega} \ L}_{ij} \ {\bf {\omega}} ~~~~~~
= \ \frac{1}{2} \ {\bf \bar{\omega}}_{ij} \ {\bf L} \ {\bf {\omega}}_{ij}
~,~~~~~ {\bf {\omega}}_{ij} \ \equiv \ {\bf {\omega}}_i - {\bf {\omega}}_j .
\end{eqnarray}

Under the group generated by the basis generator, ${\bf T}_i$,  the $J_m$ for 
instance varies as:
\begin{eqnarray}
2J_m(\sigma) & = & 2 J_m + \delta_{im} \left[ \ -sin^2(\sigma) \ \left\{
2 J_i - {\bf {\bar{\omega}_i \ (\bar{u}\ L\ u ) \ \omega_i}} \right\}
\right. \nonumber \\
& & ~~~~~~~~~~~~~~~ \left. + \ sin(\sigma) cos(\sigma) \ \left\{ \ {\bf {\bar{\omega}_i \ ( 
\bar{u} L + L u) \ \omega_i \ } } \right\} \ \right]
\end{eqnarray}

while under the group generated by the basis generator, ${\bf T}_{ij}$, 
the $J_m$ varies as:
\begin{eqnarray}
2J_m(\sigma) & = & 2 J_m + \delta_{im} \left[ \ -sin^2(\sigma) \ \left\{
2 J_i - {\bf {\bar{\omega}_j \ (\bar{v}\ L\ v ) \ \omega_j}} \right\}
\right. \nonumber \\
& & ~~~~~~~~~~~~~~~ \left. + \ 2 sin(\sigma) cos(\sigma) \ \left\{ \ {\bf {\bar{\omega}_i \ ( 
L v ) \ \omega_j \ } } \right\} \ \right] \nonumber \\
& & ~~~~~~~~~ \delta_{jm} \left[ \ -sin^2(\sigma) \ \left\{
2 J_j - {\bf {\bar{\omega}_i \ (v\ L\ \bar{v} ) \ \omega_i}} \right\}
\right. \nonumber \\
& & ~~~~~~~~~~~~~~~ \left. - \ 2 sin(\sigma) cos(\sigma) \ \left\{ \ {\bf {\bar{\omega}_i \ ( 
v L ) \ \omega_j \ } } \right\} \ \right] ~~~~~~
\end{eqnarray}

Similar but more complicated expressions follow for the $J_{mn}(\sigma)$
also. 

{\underline {Remark:}} For the case of $N = 1$ we have only one block. This 
could be either a single
two dimensional oscillator {\it or} the relative coordinate dynamics of two anyons.
The generators are of course only ${\bf u}$'s. Of these ${\bf u}_2$
corresponds to the Hamiltonian itself while ${\bf u}_1$ generates same rotations
of both $\vec{r}, \vec{p}$. These two matrices commute with {\bf L} while the
remaining two anti-commute with {\bf L}.  This leads to the result quoted in
the section 5.1 .

To deduce the surviving symmetry group for many anyons we need infinitesimal
variations of $J_{mn}$ for arbitrary generator {\bf T}.  This is easily
derived and is given by,
\begin{equation}
\delta J_{mn} \ = \ \sum_{j} {\bf {\bar{\omega}_{mn}\ L \ ( T}}_{mj} - {\bf T}_{nj} \
) {\bf {\omega}}_j 
\end{equation}

In section 5.1 we needed to determine {\bf T} such that $\delta J_{mn}$ is
zero {\it whenever} $J_{mn}$ is zero. That is, at all points in the phase
space where any single $J_{mn}$ is zero, we want its infinitesimal variation
induced by {\bf T} to be zero. 

Fix a particular $m$ and $n$. Fix ${\bf {\omega}}_{mn}$. We can consider
points with ${\bf {\omega}}_j, \ j \ne m , n$ such that no other $J_{ij}$ is
zero. $\delta J_{mn} \ = \ 0$ then implies ${\bf T}_{mj} \ = \ {\bf T}_{nj} \
\forall \ j \ne m , n$. We can also consider points with different ${\bf
{\omega}}_m , {\bf {\omega}}_n$ keeping ${\bf {\omega}}_{mn}$ fixed and
maintaining all other conditions. This implies ${\bf T}_{mm} - {\bf T}_{nm} +
{\bf T}_{mn} - {\bf T}_{nn} = 0$. Repeating this for all $m \ne n$ fixes the
form of {\bf T} in terms of arbitrary generators of $OSp(4,R)$, {\bf u} ,
{\bf v} as,
\begin{equation}
({\bf T})_{ij} \  =  \ {\bf u} \delta_{ij} \ + \ {\bf v} ( 1 - \delta_{ij} )
~~,~~ \left[ \ {\bf L} \ , \ {\bf u - v}\ \right] \  =  \ 0 
\end{equation}

The choice {\bf u} = {\bf v} corresponds to ${\bf T}_{ij} \ = \ {\bf u} ~ 
\forall \ i, j$.  It follows that {\it all} $J_{mn}$ 's are invariant independent
of their values. It is easy to see that these {\it four} {\bf T}'s
effect transformations of the center-of-mass variables which are insensitive
to the anyonic features. This $OSp(4,R)$ symmetry is thus always present for all 
$N \ge 2$. 

The choice {\bf v} = 0 implies {\bf T} is block diagonal with the same {\bf
u} on all the diagonal blocks. Further, ${\bf [ L , u ]} = 0$ implies that
{\bf u} must be a linear combination of ${\bf u}_1$ and ${\bf u}_2$ .
These {\it two} {\bf T} 's can be seen to correspond to the total Hamiltonian 
and the total angular momentum. This result is used in the section 5.1 to deduce 
that the surviving symmetry for $N$ anyons is $OSp(4,R) \times O(2,R) \times O(2,R)$.


\newpage

\subsection*{Appendix C : Symplectic Diagonalization}
 
\addcontentsline{toc}{section}{Appendix C : Sympletic Diagonalization} 

In our method of calculating semiclassical corrections to a classical
ground state, we need to diagonalise the Hessian matrix, the matrix of
second derivatives of the potential at the ground state..
In the case of anyons, since the ground state is not
static this matrix is in terms of all the
phase-space variables and not merely the coordinates.
The diagonalization must hence preserve the symplectic structure of
the phase space.
The Hamiltonian expanded about the ground state is
\beqn
H = H_0 +  \xi^T M \xi + \text{~~higher order terms},
\eeqn
where $\xi$ is the column vector made up of the phase space variables.

The phase space structure is preserved if the matrix $M$ is
diagonalised by a symplectic matrix $S$ such that
$S^T \mathbb{J} S = \mathbb{J}$,
 $\mathbb{J}$ being the $2N-2$-dimensional symplectic form.
Now there exists a theorem\footnote{ We are indebted to Prof. R.~Simon
for bringing this theorem to our notice, as well as for the simple
proof in the paper \cite{simon}} known as Williamson's theorem
\cite{will} (a compact summary is presented in Appendix 6 of Arnold
\cite{arnold}) part of which implies that if $M$ is a real symmetric
positive
definite matrix then it can be symplectically diagonalised by such an
$S$. Moreover, if $S^T M S = D^2$ where $D = \text{diag}(D_1, D_2)$,
$D_1,D_2$ being $N-1$ dimensional diagonal matrices, and
$\Tilde{M} = -\iota M^{-1/2} \mathbb{J} M^{-1/2}$ is a Hermitian
matrix with real eigenvalues $\pm \omega_i$ then it can be shown that
$D_1^{-1}D_2^{-1} = \text{diag} (\omega_i)$. When this happens, then
if the normal modes are $ {\xi}^{\prime} \equiv \bigl(
\begin{smallmatrix}
Q_i\\P_i\end{smallmatrix}\bigr) = S^{-1}\xi$, the quadratic form looks
like
\begin{equation}
\begin{split}
{\xi}^{\prime T}D^2{\xi}^{\prime} =
 &\sum_i (d_i^2 Q_i^2 + \omega_i^{-2}d_i^{-2}P_i^2)\\
          =&\omega_i^{-2}d_i^{-2} ( P_i^2 + \omega_i^2 d_i^4 Q_i^2).
\end{split}
\end{equation}
This is a harmonic oscillator in the space of normal modes, on
quantizing which we get the correction to the ground state energy as
\beqn
\delta E_0 =& \frac{1}{2} \hbar \sum_i \omega_i^{-2}d_i^{-2} \omega_i
d_i^2 \nonumber\\
   =& \frac{1}{2} \hbar \sum_i \omega_i^{-1}.
\eeqn
Thus one needs just the positive eigenvalues of the matrix
$\Tilde{M}$, which should further be non-zero. This will require $M$
itself to have non-zero eigenvalues.
 
Now to find the eigenvalues of the Hessian \eqref{Hess}:
The last row and column, coming from the derivatives with respect to $P_{X}$,
contribute an eigenvalue $1$. Consider the reduced $4N-3$ dimensional
matrix $M^{\prime}$ formed by removing the last row and the
last column. It turns out that this matrix does have zero eigenvalues,
inherited from the eigenvalues of $\Delta$ with eigenvectors of the
form $ k \bigl( \begin{smallmatrix} \zeta, & -\frac{1}{2}\zeta^T\zeta
\end{smallmatrix} \bigr)^T$, $k$ being an arbitrary proportionality
factor. There are exactly $2N-2$ of these. The remaining eigenvalues
consist of an eigenvalue equal to 2 which is $2N -2$ fold degenerate
and a non-degenerate eigenvalue equal to 
$ 2\left( 1 +\frac{|J|}{X^2} \right)$.
 

\end{document}